\documentclass[reprint,showpacs,preprintnumbers,amsmath,amssymb,prc,floatfix]{revtex4-1}
\usepackage{float}
\usepackage{color}
\usepackage{graphicx}
\usepackage{dcolumn}
\usepackage{bm}
\usepackage{xcolor}
\usepackage{multirow}
\usepackage{comment}
\usepackage[colorlinks,citecolor=blue,linkcolor=red,anchorcolor=blue,filecolor=blue,urlcolor=blue]{hyperref}

\begin{document}


\title{Fusion cross section of the superheavy $Z$ = 120 nuclei within the relativistic mean-field formalism }

\author{Shilpa Rana}
\email{srana60\_phd19@thapar.edu}
\affiliation{School of Physics and Materials Science, Thapar Institute of Engineering and Technology, Patiala 147004, India}
\author{Raj Kumar}%
\email{rajkumar@thapar.edu}
\affiliation{School of Physics and Materials Science, Thapar Institute of Engineering and Technology, Patiala 147004, India}
\author{M. Bhuyan}
\email{bunuphy@um.edu.my}
\affiliation{Center for Theoretical and Computational Physics, Department of Physics, Faculty of Science, University of Malaya, Kuala Lumpur 50603, Malaysia}
\affiliation{Institute of Research and Development, Duy Tan University, Da Nang 550000, Vietnam}

\date{\today}
\bigskip 
\begin{abstract}
\noindent
{\bf Background:} Isotopes of $Z$ = 107 - 118 have been synthesized using cold fusion at GSI, Darmstadt, and hot fusion reactions at JINR, Dubna. Recently theoretical models have predicted $Z$ = 120 with $N$ = 184 as an island of stability in the superheavy valley. Hence it is crucial and exciting to predict theoretically the possible combination of projectiles and targets for the synthesis of $Z$ = 120, which can be informative for upcoming experiments.\\ 
\noindent
{\bf Purpose:} Present theoretical investigations aim to explore the fusion characteristics of various isotopes of $Z$ = 120 within the relativistic mean-field formalism. We predict the most suitable projectile-target combination for the synthesis of element $Z$ = 120. The increase in fusion cross-section of nuclei in the superheavy island directly signals the nuclear shell effects. Besides these, the analysis will be crucial and relevant for future experiments to synthesize superheavy nuclei.  \\
\noindent
{\bf Methods:} The microscopic nucleon-nucleon R3Y interaction and the density distributions for targets and projectiles are calculated using a relativistic mean-field formalism with the NL3$^*$ parameter set. These densities and R3Y nucleon-nucleon (NN)-interaction are then used to calculate the nuclear interaction potential using the double folding approach. Seventeen different projectile-target combinations that allow a high $N$/$Z$ ratio are considered in the present analysis to calculate the capture and/or fusion cross-section of various isotopes of $Z$ = 120 within the $\ell-$summed Wong formula. \\
\noindent
{\bf Results:} The nuclear density distributions for the interacting projectile and target nuclei are obtained from relativistic mean-field Lagrangian for the NL3$^*$ parameter set. The nucleus-nucleus interaction potential is estimated for seventeen possible projectile-target combinations using the mean-field density and the R3Y NN-potential via a double folding approach. The fusion-barriers are obtained by adding the Coulomb potential to the nucleus-nucleus interaction potential. Finally, the fusion and/or capture cross-section is calculated for all the systems within the $\ell-$summed Wong formula. Further, the equivalent surface diffusion parameter is estimated to correlate the surface properties interacting nuclei with the fusion cross-section.\\
\noindent
{\bf Conclusions:} The four Ti-based reactions with the heaviest available target $^{x}$Cf, namely, $^{46}$Ti+$^{248}$Cf, $^{46}$Ti+$^{249}$Cf, $^{50}$Ti+$^{249}$Cf, and $^{50}$Ti+$^{252}$Cf, and also $^{54}$Cr+$^{250}$Cm are found to have the most suitable target-projectile combinations for the synthesis of various isotopes $Z$ = 120. We also notice that $^{48}$Ca beams merely provide the required number of protons to synthesize the element with $Z$ = 120. We established a correlation among the surface properties of interacting nuclei with the fusion characteristics in terms of the equivalent surface diffusion parameter.  \\
\end{abstract}
\pacs{21.65.Mn, 26.60.Kp, 21.65.Cd}
\maketitle

\section{INTRODUCTION}
\label{intro} \noindent
The quest to synthesize superheavy nuclei (SHN) and understand related properties is one of the central research areas in the new era of nuclear physics. With advancements in experimental laboratories, the seventh row of the Periodic Table is now complete with the latest addition of element oganesson ($Z$ = 118) \cite{ogan06}. The study of SHN is fascinating because they allow physicists to explore the concepts of magic numbers, drip-lines, and most importantly, the existence of  "island of stability" \cite{ogan12,ogan17}.  As a consequence, remarkable experimental and theoretical research is being devoted to this field \cite{ogan15,hoff15,Zagr12,myer00,nan13,bao17,arit19,kuma13}, where the main aim is not only to extend the Periodic Table but also to achieve the subsequent magic shell closure beyond $Z$ = 82 for protons and $N$ = 126 for neutrons. In this direction, various theoretical calculations have predicted the next possible proton magic number to be at $Z$ = 114 ,120 or 126 with neutron magic $N$ = 172, 182, 184, and 258 \cite{sobi66,meld67,nils69,bhuy12,adam09,paty91,myer66,hoff04,lala96,rutz97,agbe15,shi19,zhang05,tani20}. Among these studies, relativistic mean-field models with various parameter sets have concentrated on the proton magic number at $Z$ = 120 with neutron magic number $N$ = 184 \cite{bhuy12,zhang05,rutz97,agbe15,shi19}. In the present investigation, the relativistic mean-field formalism will be adopted, hence the focus of the study is to determine the fusion cross-section for the superheavy isotopes of $Z$ = 120. \\
\\
The synthesis of superheavy nuclei is a somewhat tedious task as they are highly unstable and hard to detect. The cross section of SHN is of the order of picobarn, and also they have very short half-lives (51.5 days to a few microseconds) \cite{mori07,ogan12}.
At present, heavy-ion fusion reactions are widely adopted for the synthesis of superheavy nuclei \cite{ogan15,hoff15,hoff95}. Mainly these are categorized into two groups: (a) the cold fusion reactions, which contain a strongly bound closed-shell nucleus i.e. $^{208}$Pb and $^{209}$Bi fuse with isotopes of $^x$Cr - $^x$Zn to produce a compound nucleus at low excitation energy \cite{hoff00,zagr08,hami13}; (b) the hot fusion reactions, where $^{48}$Ca beams fuse with isotopes of actinide targets to form a compound nucleus with high excitation energy \cite{ogan07,ogan212,ogan112}. Using these fusion reactions, it is possible to synthesize elements of $Z$ = 107-118 \cite{hoff00,zagr08,hami13,ogan07,ogan212,ogan112}. In the superheavy valley, the next proton closure is predicted to be $Z$ = 120 with neutron number  $N$ = 184 \cite{bhuy12,adam09,hoff04,lala96,rutz97,zhang05,agbe15,shi19}. To synthesize the isotopes of $Z$ = 120 with a $^{48}$Ca beam, a target actinide with $Z$ $ \ge$ 100 is required. These actinides are available in a limited amount to be used as targets due to very short half-lives. In 2007, Hoffman {\it et al.}, performed the reaction $^{238}$U($^{64}$Ni,$xn)^{302-x}120$ \cite{hoff08} at GSI, and reached the upper cross-section limit of 0.09 $pb$ at E$^{*}$ = 36.4 MeV. Later on, Oganessian {\it et al.} \cite{ogan09} attempted to synthesize element $Z$ = 120 in reaction  $^{244}$Pu($^{58}$Fe,$xn)^{302-x}120$, and was unable to observe any possible decay chain although the upper cross section limit was 0.4 pb. They also predicted that the cross section will be enhanced for more asymmetric projectile target combinations {\it e.g.} $^{54}$Cr+$^{248}$Cm or $^{50}$Ti+$^{249}$Cf \cite{ogan09}. In addition to these, the capture cross section for the fission like fragments of reactions $^{64}$Ni+$^{238}$U \cite{kozu10} and $^{54}$Cr+$^{248}$Cm with the upper limit of 0.56 pb \cite{hein13} was determined to synthesize $Z$ = 120. More details along with a few experimental attempts to synthesize the new element $Z$ = 120 at GSI  can be found in Refs. \cite{chris17,novi20,albe20}. \\
\\
The above experimental attempts show that one needs to adopt a feasible way; that is, to move towards a projectile with a higher proton number than that of $^{48}$Ca. Thus it will be of great significance to predict the possible combinations of target and projectile nuclei to synthesize the isotopes of $Z$ = 120. Hence, in the present study, we have considered the isotopes of calcium (Ca), titanium (Ti), chromium (Cr), iron (Fe), nickel (Ni), zinc (Zn), and germanium (Ge) as projectiles and corresponding actinides as targets for the synthesis of isotopes of $Z$ = 120. In other words, the potential barrier and the cross sections for a few target-projectile combinations for the synthesis of various isotopes of $Z$ = 120 will be investigated. In parallel to the synthesis of superheavy nuclei in the laboratory, a few theoretical calculations have also been performed for different projectile-target combinations to predict the cross-section for various isotopes of superheavy nuclei \cite{liu13,siwe12,wang12,wang12,lian12,seki19,liu06,fan18,nasi09,srid19,shar19}. In the present work, we have adopted the relativistic mean-field (RMF) approach, which has been applied successfully for the study of the ground state properties such as quadrupole deformations, binding energies, alpha decay Q$_{\alpha}$ values, fission barrier, etc. of superheavy nuclei \cite{patra09,bhuy11,sahu11,shak12,lala96, bend99, afan03, pras12, tani20,agbe15,shi19}. Hence, it will be interesting to predict the most suitable target-projectile combination for the synthesis of element $Z$ = 120 using the relativistic mean-field formalism along with $\ell$-summed Wong formula \cite{ring90,lala09,kuma09,lahi16,bhuy18,bhuy20,vret05,meng06}. Here we have applied the R3Y interaction derived from the relativistic mean-field Lagrangian to estimate the fusion cross section of different isotopes of the element with $Z$ = 120. More details on the R3Y nucleon-nucleon interaction can be found in Refs. \cite{sing12,lahi16,bhuy18,bhuy20}. The fusion cross-section for reactions  $^{40}$Ca + $^{257}$Fm, $^{48}$Ca + $^{254}$Fm, $^{46}$Ti + $^{248}$Cf, $^{46}$Ti + $^{249}$Cf, $^{50}$Ti + $^{249}$Cf, $^{50}$Ti + $^{252}$Cf, $^{50}$Cr + $^{242}$Cm, $^{54}$Cr + $^{248}$Cm, $^{58}$Fe + $^{244}$Pu, $^{64}$Ni + $^{238}$U, $^{64}$Ni + $^{235}$U, $^{66}$Ni + $^{236}$U, $^{50}$Ti + $^{254}$Cf, $^{54}$Cr + $^{250}$Cm, $^{60}$Fe + $^{244}$Pu, $^{72}$Zn + $^{232}$Th, and $^{76}$Ge + $^{228}$Ra are calculated and compared, and the most suitable projectile-target combination for the synthesis
of the element with $Z$ = 120 is also predicted. It is to be noted here that the stability of superheavy nuclei lies close to the neutron-rich side of the superheavy island, so more neutron-rich target-projectile combinations are considered in this study \cite{ogan00}. \\
\\
\noindent
This paper is structured as follows: In Sec. \ref{theory}, we briefly present the nucleon-nucleon interaction potential along with the relativistic mean-field approach. The $\ell$-summed Wong formula is also included in this section. Further Sec. \ref{results} shows the results for cross sections and also related physical quantities. Finally, in Sec. \ref{summary}, we discuss the conclusions and perspectives of the present study in terms of fusion cross-section of superheavy nuclei of $Z$ = 120.\\
\vspace{-0.5cm}
\section{Nuclear Interaction Potential from relativistic mean-field formalism}
\label{theory}
\noindent
The interaction potential barrier arises from the competition between the long-range repulsive Coulomb and the short-range attractive nuclear interaction. To estimate the barrier characteristics, such as height and width, one needs to calculate the total interaction potential [$V_T^{\ell}(R)$] between the target and projectile nuclei, which is given as
 \begin{eqnarray}
V_{T}^{\ell}(R)= V_n (R)+V_C(R)+ V_{\ell}(R).
\label{vtot}
\end{eqnarray}
Here, $V_C(R)$ is the Coulomb potential given by, $V_C(R)=Z_pZ_te^2/R$, and $V_{\ell}(R)$ is the centrifugal potential given by, $V_{\ell}(R)=\frac{\hbar^2\ell(\ell+1)}{2\mu R^2}$.
$V_n(R)$ is the nuclear potential which is calculated here using the double folding procedure \cite {satc79}
\begin{eqnarray}
V_{n}(\vec{R}) & = &\int\rho_{p}(\vec{r}_p)\rho_{t}(\vec{r}_t)V_{eff}
\left( |\vec{r}_p-\vec{r}_t +\vec{R}| {\equiv}r \right) \nonumber \\
&& d^{3}r_pd^{3}r_t, 
\label{fold}
\end{eqnarray}
where $\rho_p$ and $\rho_t$ are densities of projectile and target nuclei, respectively. $V_{eff}$ is the effective nucleon-nucleon (NN) interaction. The densities and effective NN interaction are calculated from the well-known relativistic mean-field (RMF) formalism, which has been used successfully to describe the properties of infinite nuclear matter, as well as the finite nuclei over the nuclear chart, including the exotic and superheavy nuclei \cite{patra09,bhuy11,sahu11,shak12,lala09,ring96,vret05,meng06,meng16}. In RMF theory, the nucleons are considered as pointlike particles denoted by Dirac spinors $\psi$ interacting through the exchange of effective pointlike particles: mesons and photons. The phenomenological description of the nucleon-meson many-body system can be given by a Lagrangian density of the form \cite{ring96,bogu77,rein89,sero86,lala09,bhuy18,bhuy20,meng06,meng16}
\begin{eqnarray}
{\cal L}&=&\overline{\psi}\{i\gamma^{\mu}\partial_{\mu}-M\}\psi +{\frac12}\partial^{\mu}\sigma
\partial_{\mu}\sigma \nonumber \\
&& -{\frac12}m_{\sigma}^{2}\sigma^{2}-{\frac13}g_{2}\sigma^{3} -{\frac14}g_{3}\sigma^{4}
-g_{s}\overline{\psi}\psi\sigma \nonumber \\
&& -{\frac14}\Omega^{\mu\nu}\Omega_{\mu\nu}+{\frac12}m_{w}^{2}\omega^{\mu}\omega_{\mu}
-g_{w}\overline\psi\gamma^{\mu}\psi\omega_{\mu} \nonumber \\
&&-{\frac14}\vec{B}^{\mu\nu}.\vec{B}_{\mu\nu}+\frac{1}{2}m_{\rho}^2
\vec{\rho}^{\mu}.\vec{\rho}_{\mu} -g_{\rho}\overline{\psi}\gamma^{\mu}
\vec{\tau}\psi\cdot\vec{\rho}^{\mu}\nonumber \\
&&-{\frac14}F^{\mu\nu}F_{\mu\nu}-e\overline{\psi} \gamma^{\mu}
\frac{\left(1-\tau_{3}\right)}{2}\psi A_{\mu}.
\label{lag}
\end{eqnarray}
Here the masses of nucleon, $\sigma$, $\omega$, and $\rho$ mesons are denoted as M, $m_{\sigma}$, $m_{\omega}$, and $m_{\rho}$, respectively. $g_{\sigma}$, $g_{\omega}$, $g_{\rho}$ denote linear coupling constants for respective mesons. The constants $g_2$ and $g_3$ represent the self-interactions of the non-linear $\sigma-$meson field, which take care of saturation properties by generating a long range repulsive NN potential.  $A_{\mu}$, $\tau$ and $\tau_3$, denote electromagnetic field, isospin, and its third component, respectively. $F^{\mu\nu}$, $\Omega^{\mu\nu}$, and $\vec{B}^{\mu\nu}$ denote the vector field tensors for the $\omega^{\mu}$, $\vec{\rho}_{\mu}$ and photon, respectively and are given as \cite{lala09}
 \begin{eqnarray}
F^{\mu\nu} =\partial_{\mu} A_{\nu}-\partial_{\nu} A_{\mu}  \\
\Omega_{\mu\nu} = \partial_{\mu} \omega_{\nu} - \partial_{\nu} \omega_{\mu}
\end{eqnarray}
and
\begin{eqnarray}
\vec{B}^{\mu \nu}=\partial_{\mu} \vec{\rho}_{\nu} -\partial_{\nu} \vec{\rho}_{\mu}.
\end{eqnarray}
The equations of motion for nucleon and meson fields are obtained from the Lagrangian density in Eq.(\ref{lag}) and are given as
\begin{eqnarray}
&& \Bigl(-i\alpha.\bigtriangledown+\beta(M+g_{\sigma}\sigma)+g_{\omega}\omega+g_{\rho}{\tau}_3{\rho}_3 \Bigr){\psi} = {\epsilon}{\psi}, \nonumber \\
&& \left( -\bigtriangledown^{2}+m_{\sigma}^{2}\right) \sigma (r)=-g_{\sigma}{\rho}_s(r)-g_2\sigma^2 (r) - g_3 \sigma^3 (r),\nonumber  \\ 
&& \left( -\bigtriangledown^{2}+m_{\omega}^{2}\right) \omega (r)=g_{\omega}{\rho}(r),\nonumber   \\  
&& \left( -\bigtriangledown^{2}+m_{\rho}^{2}\right) \rho (r)=g_{\rho}{\rho}_3(r).
\end{eqnarray}
The nonlinear parts of the scalar meson $\sigma$ proportional to $\sigma^3$ and $\sigma^4$ in the above equation describe the self-coupling among the $\sigma$ mesons and are adjusted to the surface properties of finite nuclei. \\
The nuclear interaction potential mainly contains two parts: (i) the hardcore repulsive part with short range ($\le$ 0.4 fm) and (ii) the softcore attractive part with intermediate range. The isoscalar $\sigma$ meson generates the most dominating feature of the attractive nuclear interaction, whereas the isoscalar vector $\omega$ meson account for the hardcore repulsion of the atomic potential. The isovector vector $\rho$ meson contributes to a highly repulsive core near the center and accounts for attractive behavior near the intermediate range. Thus solving the field equations for mesons in the limit of one-meson exchange, the resultant effective R3Y NN potential $V_{eff}^{R3Y}$ \cite{bhuy20,bhuy18,sing12,lahi16} can be given by
\begin{eqnarray}
V_{eff}^{R3Y}(r)=\frac{g_{\omega}^{2}}{4{\pi}}\frac{e^{-m_{\omega}r}}{r}
+\frac{g_{\rho}^{2}}{4{\pi}}\frac{e^{-m_{\rho}r}}{r}
-\frac{g_{\sigma}^{2}}{4{\pi}}\frac{e^{-m_{\sigma}r}}{r} \nonumber \\
+\frac{g_{2}^{2}}{4{\pi}} r e^{-2m_{\sigma}r}
+\frac{g_{3}^{2}}{4{\pi}}\frac{e^{-3m_{\sigma}r}}{r}
+J_{00}(E)\delta(r) \nonumber \\
\label{r3y}
\end{eqnarray}
The last term is a pseudopotential which includes the effects of single nucleon exchange. The parameters set NL3$^*$ \cite{lala09} is used to obtain the $V_{eff}^{R3Y}$ in terms of masses and coupling constants of the mesons. Using the effective R3Y NN interaction and densities, which are also obtained from the RMF Lagrangian in Eqn. (\ref{fold}), we calculate the nuclear/nucleus-nucleus interaction potential. The total nucleus-nucleus interaction potential is then employed to obtain the capture and/or fusion cross-section from the $\ell$-summed Wong formula and the compound nucleus formation probability discussed in detail upcoming subsections. It is to be noted here that the pairing has a crucial role in the bulk properties, including the density distributions of open-shell nuclei in their ground and intrinsic excited states, see \cite{kara10} and reference therein. At present, the pairing correlation is taken care of by a few widely used methods in the mean-field model, such as the BCS approach, the Bogoliubov transformation, and the particle number conserving methods \cite{zeng83,moli97,zhang11,hao12,lala99,lala99a,bhuy18}. In principle, the BCS pairing is not suitable to deal with the drip-line nuclei. In the present study, we are dealing with the nuclei near the $\beta$-stable region of the nuclear chart \cite{zeng83,moli97,zhang11,hao12,lala99,lala99a,bhuy18}. Hence the BCS pairing approach is suitable and simpler to reproduce the reasonable paring correlation \cite{lala99,lala99a,bhuy18}. The blocking procedure is used to take care of the odd-mass nuclei. One can find more detail on the blocking procedure in Refs. \cite{bhuy18,doba84,madl88}.
\subsection{Capture cross-section from $\ell$-summed Wong formula}
\noindent
The well-known $\ell$-summed Wong formula \cite{kuma09}, which is the extended form of the Wong formula \cite{wong73}, given by Gupta and collaborators \cite{kuma09} in terms of summation over the $\ell$ partial wave, is used to study the fusion and/or capture cross section. The capture cross section in terms of the partial wave is given by
\begin{eqnarray}
\sigma(E_{c.m.})=\frac{\pi}{k^{2}} \sum_{\ell=0}^{\ell_{max}}(2\ell+1)P_{\ell}(E_{c.m}).
\label{crs}
\end{eqnarray}
Here, E$_{c.m.}$ is the center-of-mass energy of two spherical colliding nuclei and $k=\sqrt{\frac{2 \mu E_{c.m.}}{\hbar^{2}}}$. The $\ell_{max}$-values are obtained from the sharp cut-off model \cite{beck81}. The quantity $\mu$ is the reduced mass and $P_{\ell}$ is known as the transmission coefficient of the total interaction potential for $\ell$th partial wave [see Eq. (\ref{vtot})] which can be determined by using the Hill-Wheeler approximation \cite{hill53}. In terms of barrier height $V_{B}^{\ell}$ and curvature $\hbar\omega_{\ell}$, $P_{\ell}$ is written as
\begin{eqnarray}
 P_{\ell}=\Bigg[1+exp\bigg(\frac{2 \pi V_{B}^{\ell}-E_{c.m.}}{\hbar \omega_{\ell}}\bigg)\bigg]. \nonumber\\
 \end{eqnarray}
The $\hbar \omega_{\ell}$ is evaluated at the barrier position $R = R_{B}^{\ell}$ corresponding to the barrier height $V_{B}^{\ell}$, and is given as
\begin{eqnarray}
 \hbar \omega_{\ell}=\hbar [|d^2V_{T}^{\ell}(R)/dR^2|_{R=R_{B}^{\ell}}/\mu]^{\frac{1}{2}}.
 \end{eqnarray}
 \\
The barrier position $R = R_{B}^{\ell}$ is obtained from the condition,
 \begin{eqnarray}
 |dV_{T}^{\ell}/dR|_{R=R_{B}^{\ell}}=0.
 \end{eqnarray}
Wong \cite{wong73} carried out the $\ell$ summation in Eq. (\ref{crs}) under approximations
(i) $\hbar \omega_{\ell}\approx \hbar \omega_{0}$, and (ii) $V_{B}^{\ell}\approx V_{B}^{0}+\frac{\hbar^2\ell(\ell+1)}{2\mu {R_{B}^{0}}^2}$, assuming $R_{B}^{\ell}\approx R_{B}^{0}$. \\
Using these two approximations and replacing the $\ell$ summation in Eq. (\ref{crs}) by integration gives the $\ell=0$ barrier-based simple Wong formula \cite{wong73} of the form
\begin{eqnarray}
\sigma(E_{c.m.}) &=& \frac{{R_{B}^{0}}^{2}\hbar\omega_{0}}{2E_{c.m.}} \nonumber \\ 
&& ln\bigg[1 + exp\bigg(\frac{2\pi}{\hbar\omega_{0}}(E_{c.m.}-V_{B}^{0})\bigg)\bigg]. 
\end{eqnarray}
This is a simplified formula to calculate the capture cross-section for two spherical colliding nuclei using the barrier characteristics $V_{B}^{0}$, $R_{B}^{0}$, and $\hbar \omega_{0}$ within the barrier penetration model. However, the $\ell$-summation procedure introduced by Wong using only the $\ell = 0$ barrier excludes the actual modifications entering the potential due to its angular momentum dependence. Details can be found in Refs. \cite{kuma09,bhuy18}. In the present work, a more precise and accurate formula is given in Eqn. (\ref{crs}), which includes the actual $\ell$ dependence of interaction potential, and is used for calculating the capture cross section.\\
The $\ell-$summed Wong formula described above gives the fusion cross sections for light, medium, and heavy nuclei. But since we are doing the calculations for $Z$ = 120, a superheavy nucleus, the probability of compound nucleus formation must be considered. For the superheavy mass region, the probability of a compound nucleus decreases with an increase in the atomic mass. Thus the fusion cross section for a superheavy nucleus is given as
\begin{eqnarray}
\sigma_{fus}= \sigma \times P_{CN}.
\label{fus}
\end{eqnarray}
Here, the $P_{CN}$ is probability of formation of a completely fused compound nucleus after the capture stage. It is an energy-dependent function given as \cite{love07,zagr08,armb99,adam00}
\begin{eqnarray}
P_{CN}= \frac{exp[-c(X_{eff}-X_{thr})]}{1+exp(\frac{V^{*}_{B}-E^*}{\Delta})}.
\label{pcn}
\end{eqnarray}
Here $\Delta\approx 4$  MeV is an adjustable parameter and $E^*$ is compound nucleus excitation energy. $V^{*}_{B}$ is the excitation energy of the compound nucleus at $E_{c.m.}\approx$ Coulomb barrier and $X_{eff}$ is effective fissility given as
\begin{eqnarray}
X_{eff}=\bigg[\frac{Z^2/A}{(Z^2/A)_{crit}}\bigg][1-\alpha+\alpha f(k)],
\end{eqnarray}
with
\begin{eqnarray}
&& \bigg(Z^2/A\bigg)_{crit}=50.883\bigg[1-1.7826 \bigg(\frac{(N-Z)}{A}\bigg)^2\bigg], \\
\nonumber \\
&& f(k)=\frac{4}{k^2+k+\frac{1}{k}+\frac{1}{k^2}}, 
\end{eqnarray}
and
\begin{eqnarray}
 && k=\bigg(A_1/A_2\bigg)^{\frac{1}{3}}.
\end{eqnarray}
Here, $Z$, $N$, and $A$ are proton, neutron, and mass number of the compound nucleus, respectively. $A_1$ and $A_2$ denote the mass numbers of projectile and target, respectively. More detailed expressions and fitting parameters for hot and cold fusion can be found in Ref. \cite{love07}.

\begin{figure*}
\centering
\includegraphics[scale=0.22]{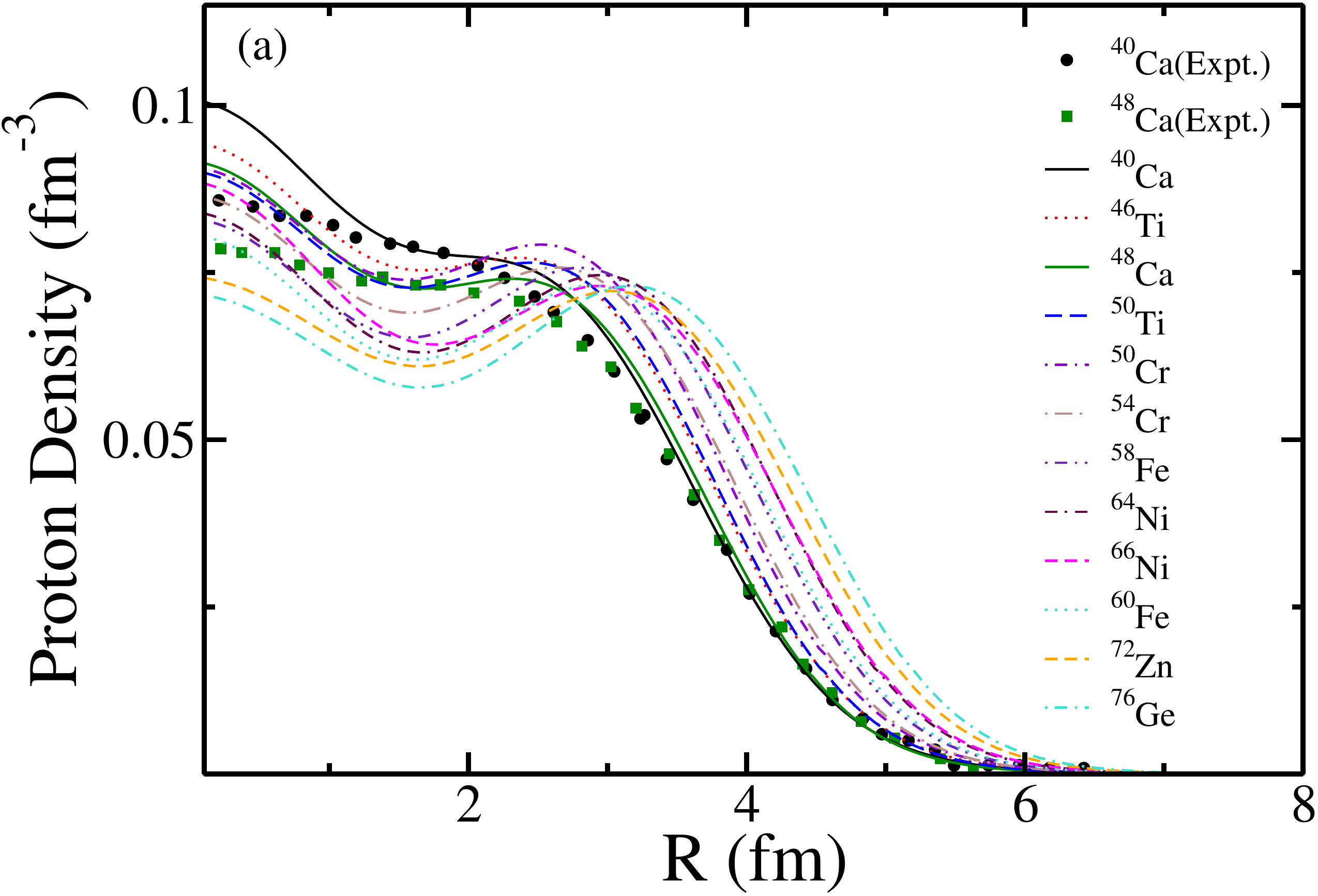}
\includegraphics[scale=0.22]{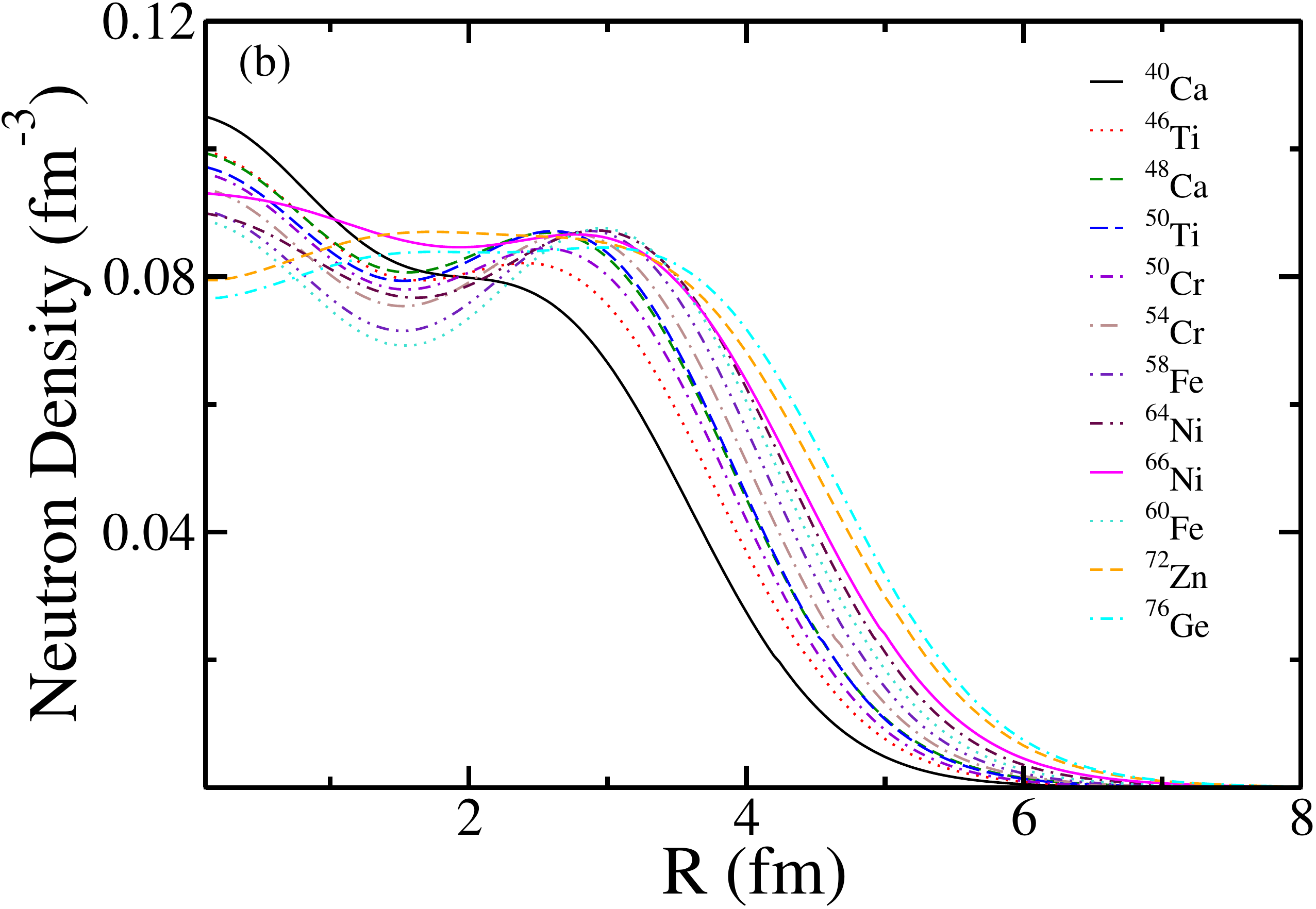}
\includegraphics[scale=0.22]{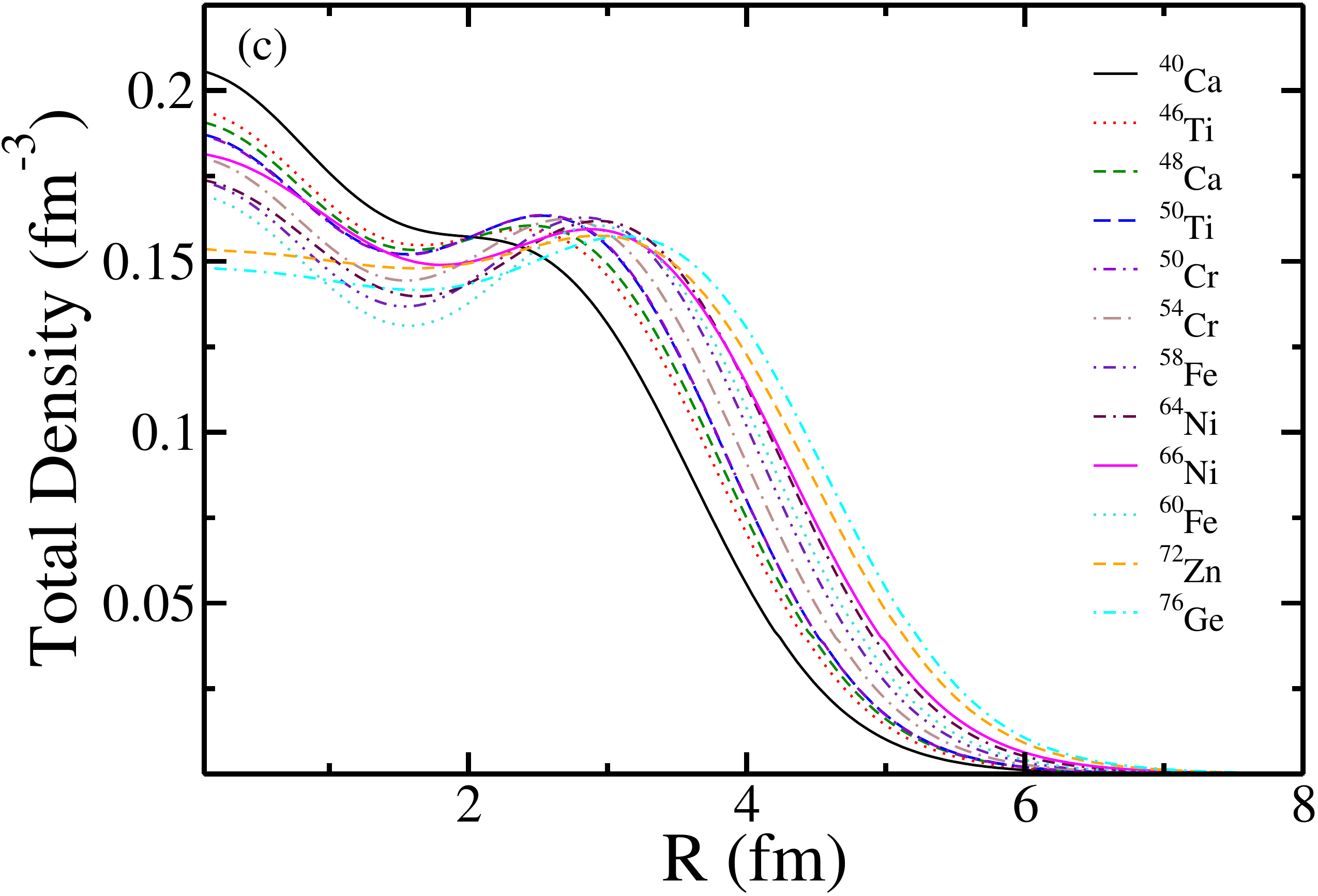}
\includegraphics[scale=0.22]{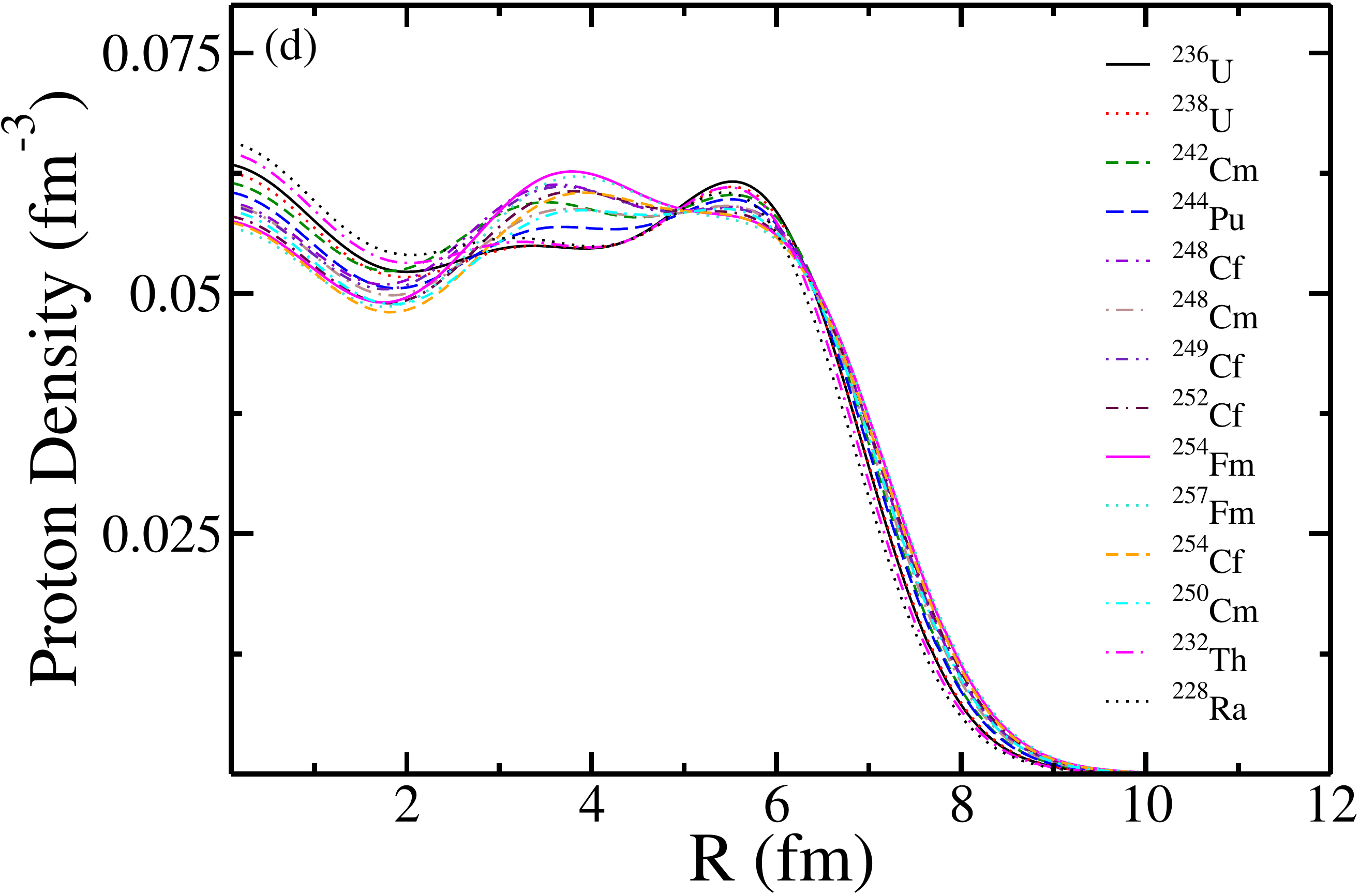}
\includegraphics[scale=0.22]{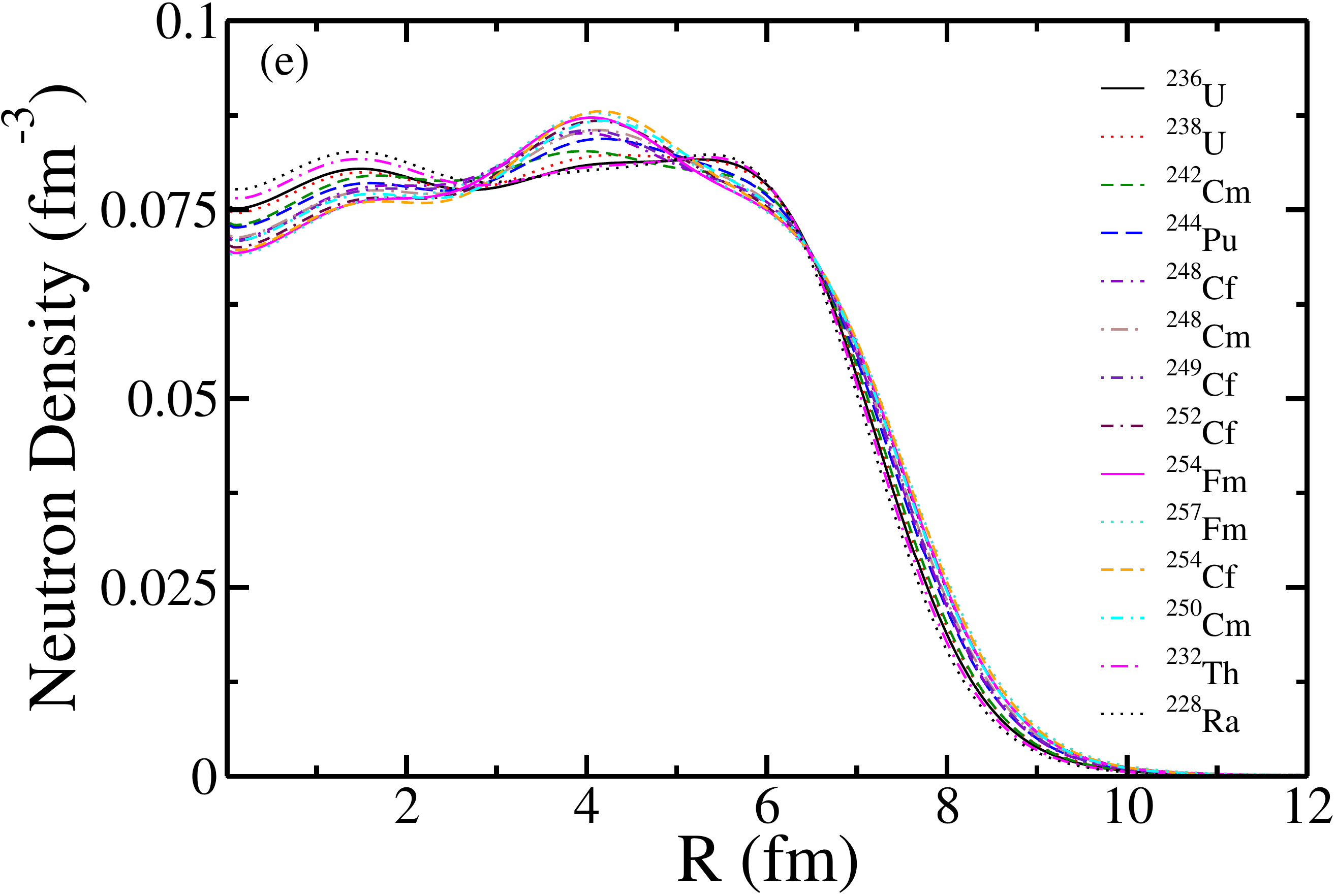}
\includegraphics[scale=0.22]{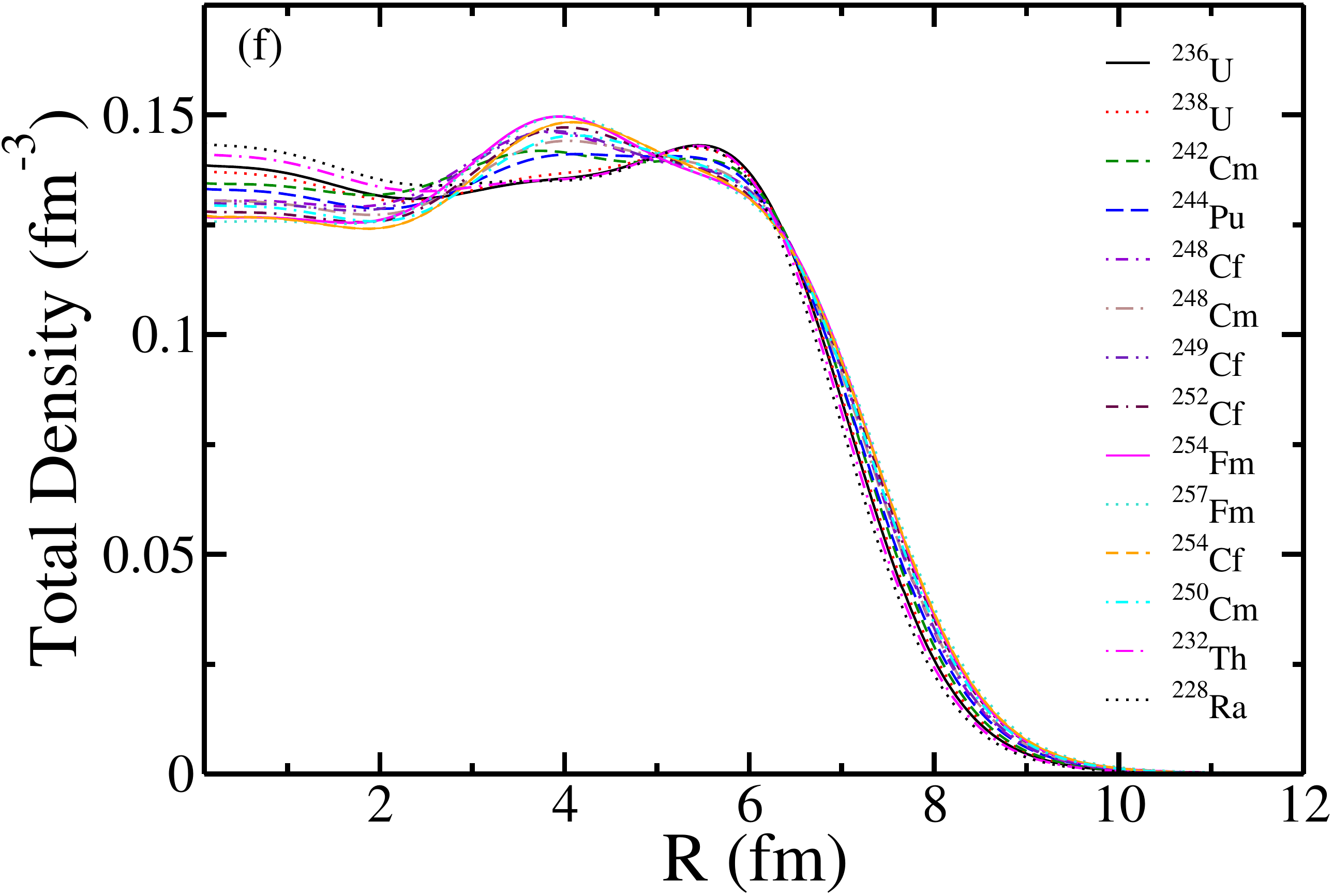}
\caption{\label{fig1} (Color online) The RMF (NL3*) proton, neutron, and total (from left to right) radial density distributions for all the considered interacting projectile (upper panel) and target (lower panel) nuclei.  See the text for details.}
\label{fig1}
\end{figure*}
\begin{figure}
\centering
\includegraphics[scale=0.33]{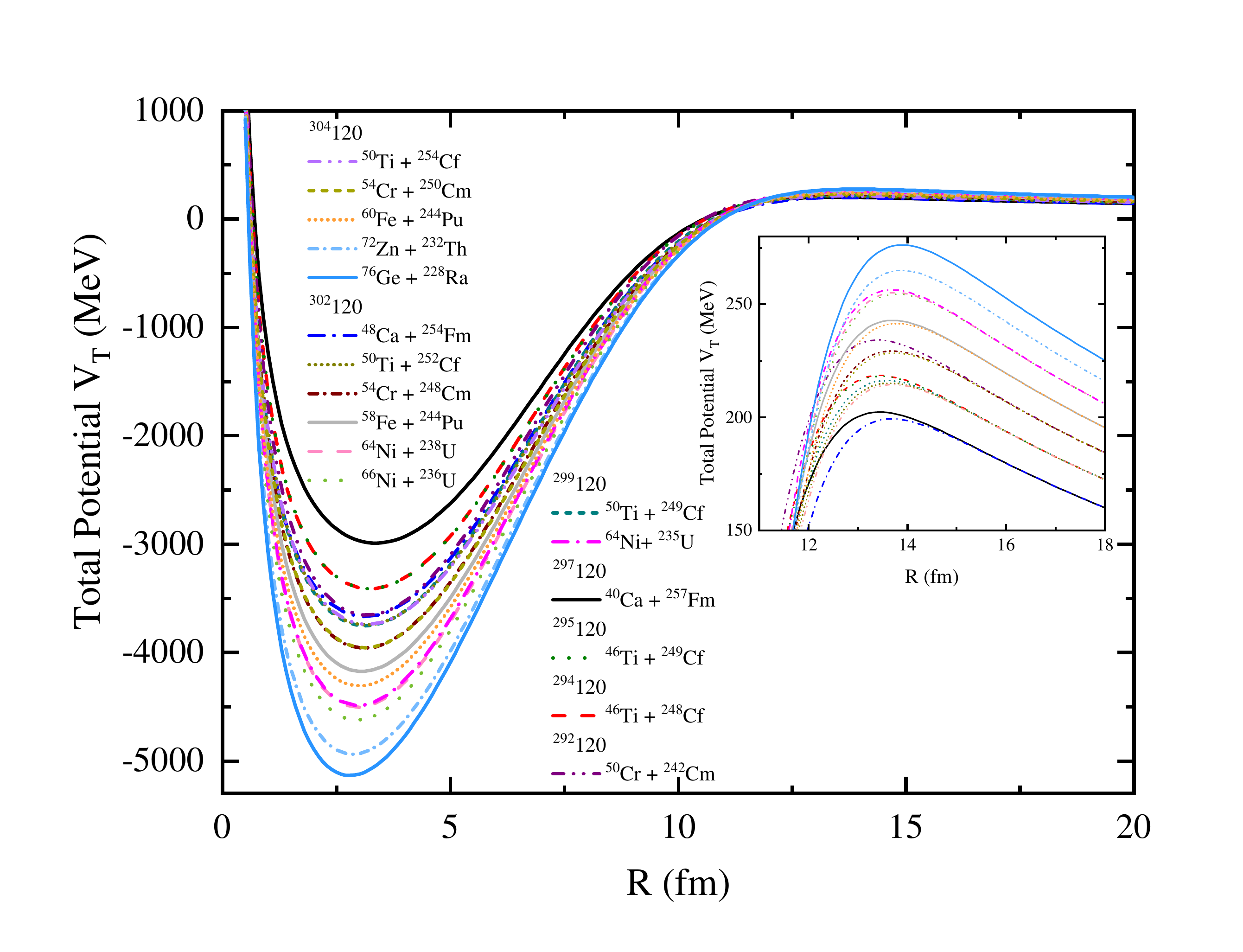}
\caption{(Color online) The total interaction potential [$V_T(R) = V_n(R) + V_C(R)$] at $\ell=0$ as a function of radial separation $R$ for the relativistic mean-field formalism using the NL3$^*$ parameter set  for all the target-projectile systems considered. In the gray-scale version, the upper (lower), middle and lower (upper) solid lines in the main (inset) figure denote $^{40}$Ca+ $^{257}$Fm, $^{58}$Fe+ $^{244}$Pu and $^{76}$Ge+ $^{228}$Ra systems, respectively.}
\label{fig2}
\end{figure}
\begin{figure*}
\includegraphics[scale=.45]{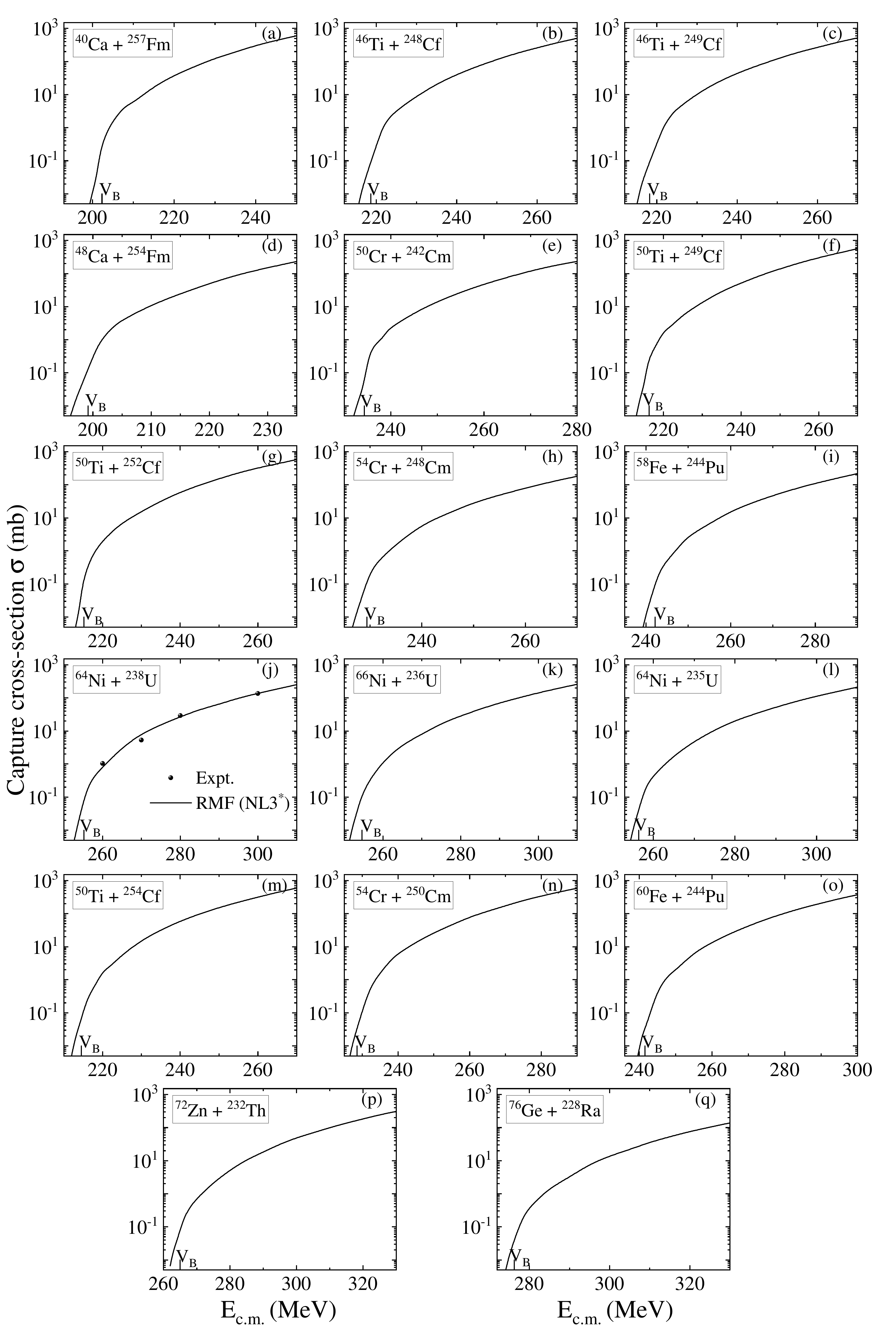}
\vspace{-0.5cm}
\caption{The capture cross section ($\sigma$) as a function of center-of-mass energy  $E_{c.m.}$ for all target-projectile systems calculated using the $\ell-$summed Wong formula. The experimental cross section for reaction $^{64}$Ni + $^{238}$U \cite{kozu10} is given for comparison. The calculated fusion barrier height ($V_B$) is also indicated for each reaction in its respective panel.}
\label{fig3}
\end{figure*}
\begin{figure}
\centering
\includegraphics[scale=0.36]{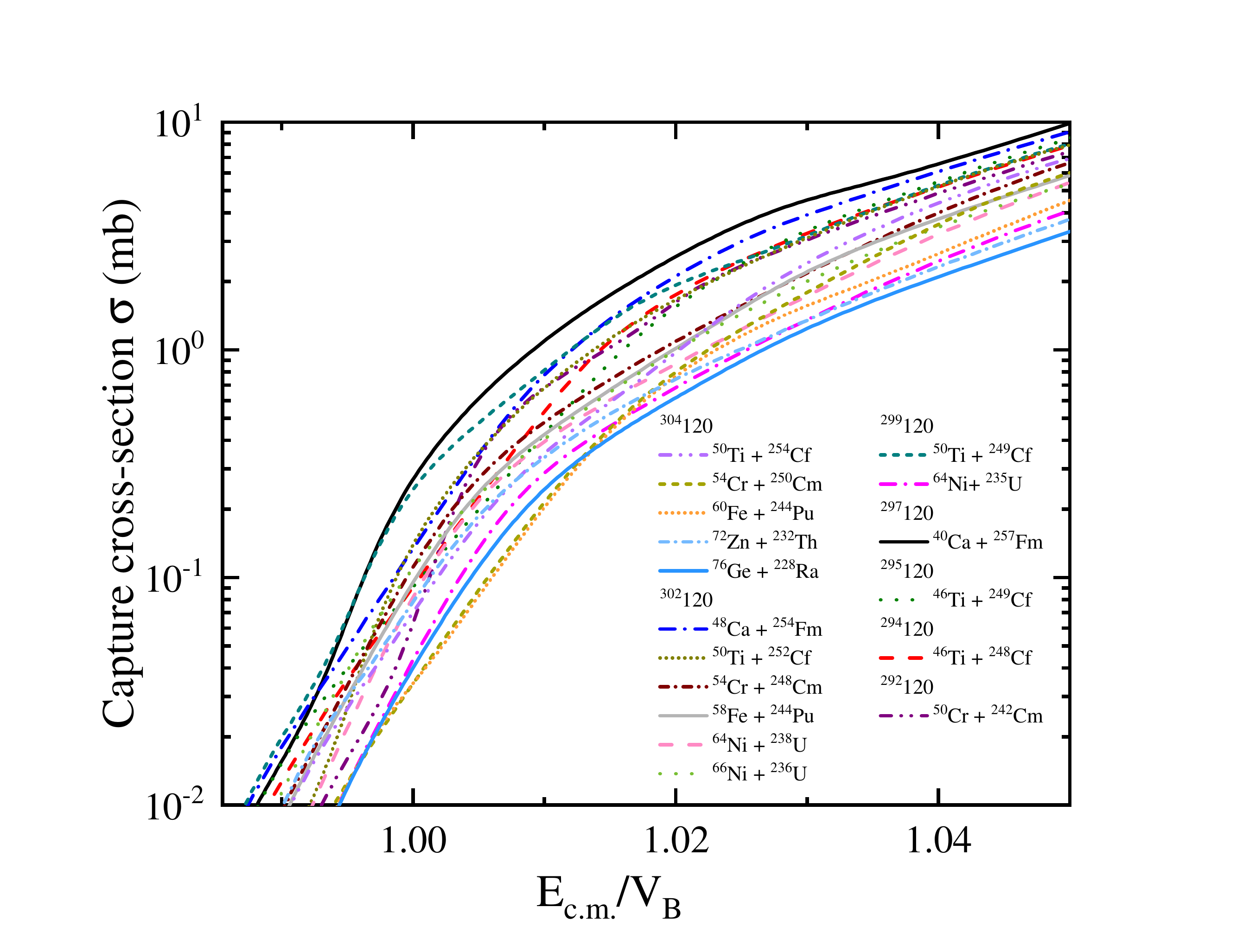}
\caption{(Color online) Comparison of capture cross section ($\sigma$) as a function of center-of-mass energy  $E_{c.m.}$ divided by estimated fusion barrier $V_B$ for all target-projectile systems calculated using the $\ell-$summed Wong formula. In the grayscale version, the upper, middle, and lower solid lines denote $^{40}$Ca+ $^{257}$Fm, $^{58}$Fe+ $^{244}$Pu, and $^{76}$Ge+ $^{228}$Ra systems, respectively.}
\label{fig4}
\end{figure}

\section{CALCULATIONS AND DISCUSSION}
\label{results}
\noindent
The fusion dynamics of superheavy nuclei involves three stages: first, the projectile overcomes/penetrates the fusion barrier formed due to Coulomb and nuclear interactions between the colliding nuclei. In the second stage, the projectile is captured by the target and forms a compound nucleus (CN). Finally, the CN system de-excites through the emission of $\gamma$ rays, neutrons, protons, and a light $N$ = $Z$ nucleus. Here in the present analysis, fusion involves forming a compound nucleus and deexciting through $\gamma$ energy. The calculation is associated with three distinct steps: (1) the nuclear density distributions of interacting nuclei are obtained using the microscopic relativistic mean-field model; (2) the nuclear interaction potential is estimated with the double folding method by using the relativistic R3Y NN potential; (3) and finally, the total interaction potentials are used to obtain the capture and/or fusion cross section and related physical quantities by adopting the $\ell$-summed Wong formula. More details of these steps are elaborated in Sec. \ref{theory}.       
 
\subsection{RMF Density distributions and Interaction Potential} 
\noindent
The interaction characteristics of two colliding nuclei can be determined through the nucleus-nucleus (in short nuclear) interaction potential. Equation (\ref{vtot}) represents the analytical expression for the total interaction potential. The densities and the relativistic R3Y nucleon-nucleon interaction involved in Eq. (\ref{vtot}) are obtained from the relativistic mean-field Lagrangian for the NL3$^*$ parameter set. Figure \ref{fig1} shows the proton, neutron, and total (from left to right) radial density distributions for all the interacting projectile (upper panel) and target (lower panel) nuclei. The experimental charge density distributions as a function of radius taken from ref. \cite{ridh17} for projectiles $^{40}$Ca (black circles) and $^{48}$Ca (green squares) are also presented in Fig. \ref{fig1} for comparison. We find a reasonably good agreement of calculated proton densities and experimental charge densities at $r > 3$ $fm$. A small discrepancy between the RMF proton densities and the experimental charge densities is observed at lower radial separation ($r<1$ $fm$). It is worth mentioning that the densities are obtained from RMF for protons without accounting for the finite size effect. Hence, there is a small difference in the central density. However, the density distributions at the surface/tail region are mainly significant for heavy-ion collisions \cite{raj07}. It can be observed from Fig. \ref{fig1} that the magnitude of densities of light mass projectile nuclei shows a small smooth dip before reaching the surface region. The combined effects from Coulomb repulsion and shell correction cause this small discrepancy at the central density, and more details can be found in Refs. \cite{rein02,afan05,chu10}.

The well-known double folding approach in which an effective nucleon-nucleon interaction is averaged over the matter density distributions is employed to calculate the nuclear interaction potential for the fusing nuclei. The relativistic R3Y NN potential and nuclear density distributions for the NL3$^*$ parameter set estimate the nuclear interaction potential via a double folding procedure.  More details of the relativistic nucleon-nucleon potential, a nuclear potential using a double folding approach, and its applicability in various studies can be found in Refs. \cite{satc79,sing12} and references therein. The total interaction potential [see Eq. \ref{vtot}] is presented in Fig. \ref{fig2} as a function of radial separation (R) at $\ell=0$ for the seventeen different target-projectile combinations considered in the present study i.e. $^{40}$Ca + $^{257}$Fm, $^{48}$Ca + $^{254}$Fm, $^{46}$Ti + $^{248}$Cf, $^{46}$Ti + $^{249}$Cf, $^{50}$Ti + $^{249}$Cf, $^{50}$Ti + $^{252}$Cf, $^{50}$Cr + $^{242}$Cm, $^{54}$Cr + $^{248}$Cm, $^{58}$Fe + $^{244}$Pu, $^{64}$Ni + $^{238}$U, $^{64}$Ni + $^{235}$U, $^{66}$Ni + $^{236}$U, $^{50}$Ti + $^{254}$Cf, $^{54}$Cr + $^{250}$Cm, $^{60}$Fe + $^{244}$Pu, $^{72}$Zn + $^{232}$Th, and $^{76}$Ge + $^{228}$Ra.
It is worth mentioning that in each combination of the considered projectiles and targets, either contain one or both of the interacting nuclei have higher $N$/$Z$ ratio i.e.; it is an isospin asymmetric system. This allows reaching the neutron-rich side of the superheavy landscape, the expected stability region of the superheavy island; see \cite{ogan00,bhuy12,shi19,sama21,zhang05} and reference therein.

The inset figure shows the barrier region of the interaction potential. Comparing the plots for all the systems, it can be noticed that the systems $^{76}$Ge + $^{228}$Ra and $^{40}$Ca + $^{257}$Fm have, respectively, the deepest and the shallowest potential pockets among the considered systems.  Further, there is an increase in the potential pocket depth with an increase in the mass number of the projectile nucleus. Comparing the height of the fusion barrier for all the systems (in the inset figure) we find that the systems $^{76}$Ge + $^{228}$Ra and $^{48}$Ca + $^{254}$Fm have the largest and lowest fusion barriers, respectively. The height of the fusion barrier shows an increase with the atomic number ($Z$) of the projectile nucleus. For isotopes of the same projectile nucleus, the fusion-barrier's height increases with the decrease in their mass number. Comparing the barrier heights of the systems leading to the formation of isotopes $^{302}120$ and $^{304}120$, it is observed that the $V_B$ increase with the decrease in $A_2/A_1$ ratio ($A_2$ and $A_1$ are masses of target and projectile, respectively). All these observations imply that a combination of light mass projectile and heavy mass target (larger $A_2/A_1$ ratio) would be the preferable better choice for forming compound nuclei with $Z$ = 120.

\begin{figure}
\centering
\includegraphics[scale=0.36]{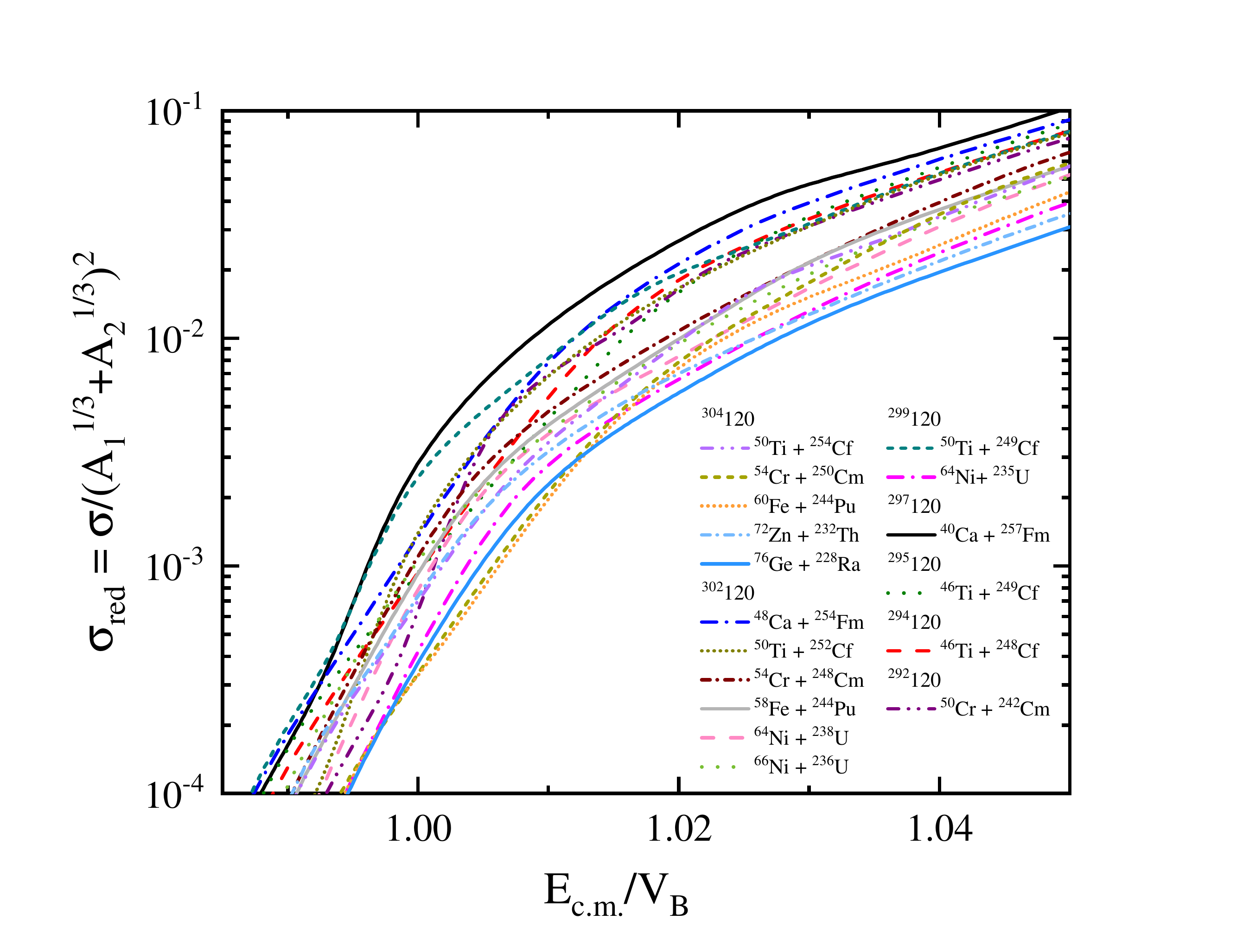}
\caption{(Color online) Capture cross section divided by the square of the interaction radius versus the center-of-mass energy  $E_{c.m.}$ divided by estimated fusion barrier $V_B$ for all target-projectile systems. In the grayscale version, the upper, middle and lower solid lines denote $^{40}$Ca+ $^{257}$Fm, $^{58}$Fe+ $^{244}$Pu and $^{76}$Ge+ $^{228}$Ra systems, respectively.}
\label{fig5}
\end{figure}
\begin{figure}
\centering
\vspace{-0.8cm}
\includegraphics[scale=0.35]{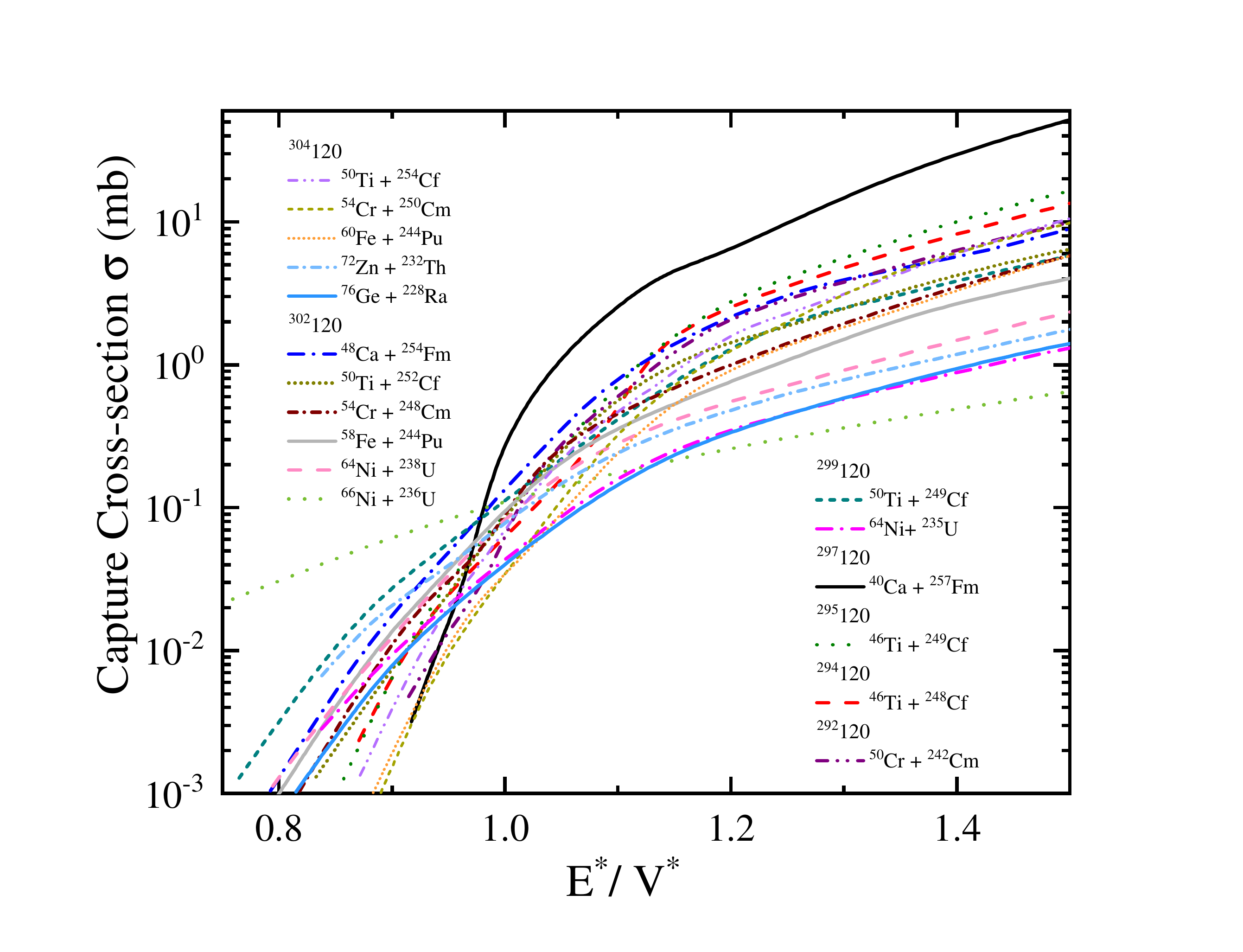}
\vspace{-0.8cm}
\caption{(Color online) The capture cross-section ($\sigma$) vs $E^*/V^*$ ($E^*=E_{c.m.} + Q_{IN}$ and $V^*=V_{B} + Q_{IN}$) for all the target-projectile combinations leading to different isotopes of SHN $Z$ = 120. In the grayscale version, the upper, middle and lower solid lines denote $^{40}$Ca+ $^{257}$Fm, $^{58}$Fe+ $^{244}$Pu and $^{76}$Ge+ $^{228}$Ra systems, respectively.} 
\label{fig6}
\end{figure}
\begin{figure}
\centering
\includegraphics[scale=0.35]{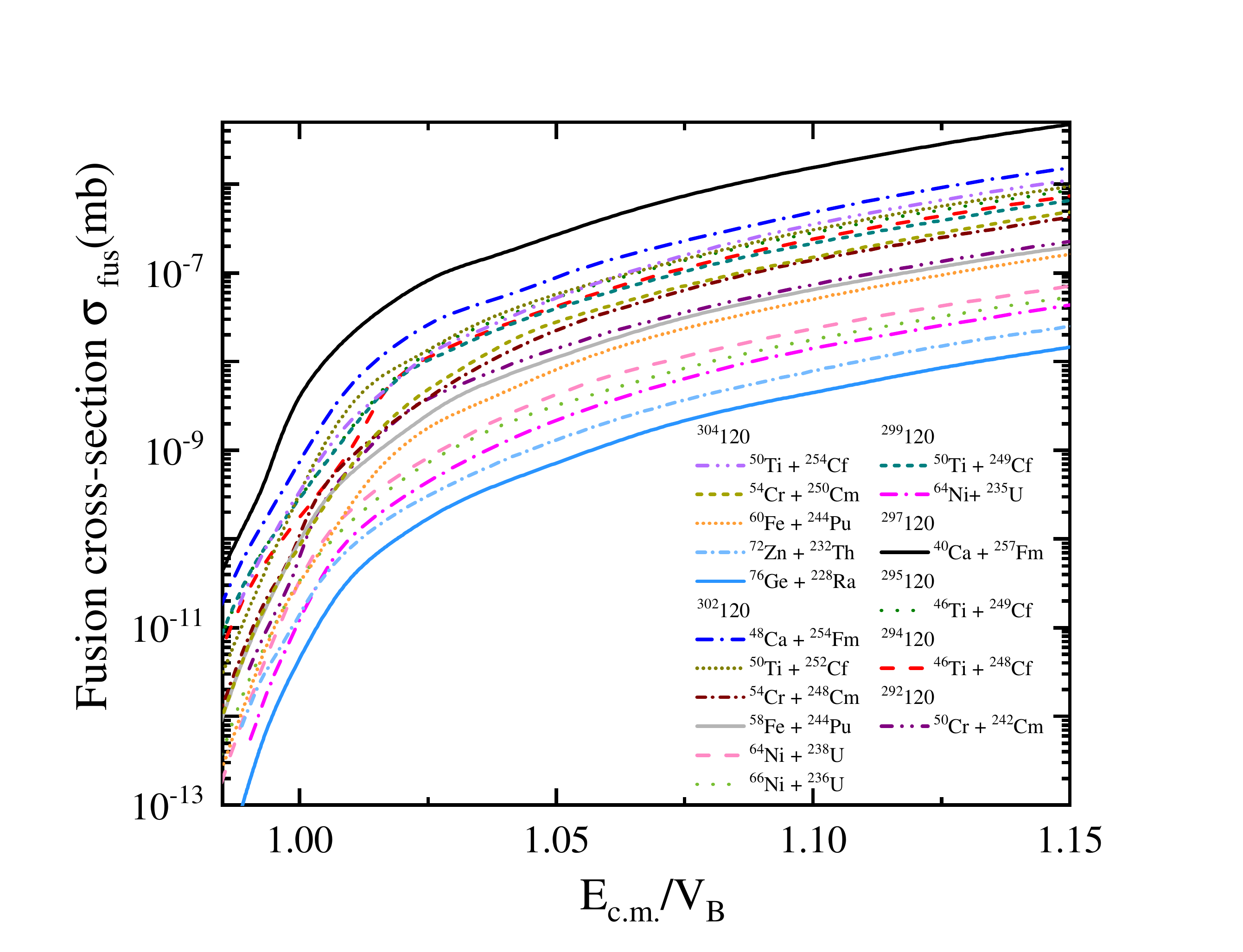}
\caption{(Color online) The fusion cross section $\sigma_{fus}$ (mb) as a function of center-of-mass energy  $E_{c.m.}$  is divided by estimated fusion barrier $V_B$ for all target-projectile systems. In the grayscale version, the upper, middle and lower solid lines denote $^{40}$Ca+ $^{257}$Fm, $^{58}$Fe+ $^{244}$Pu and $^{76}$Ge+ $^{228}$Ra systems, respectively.}
\label{fig7}
\end{figure}
\begin{figure}
\centering
\includegraphics[scale=0.37]{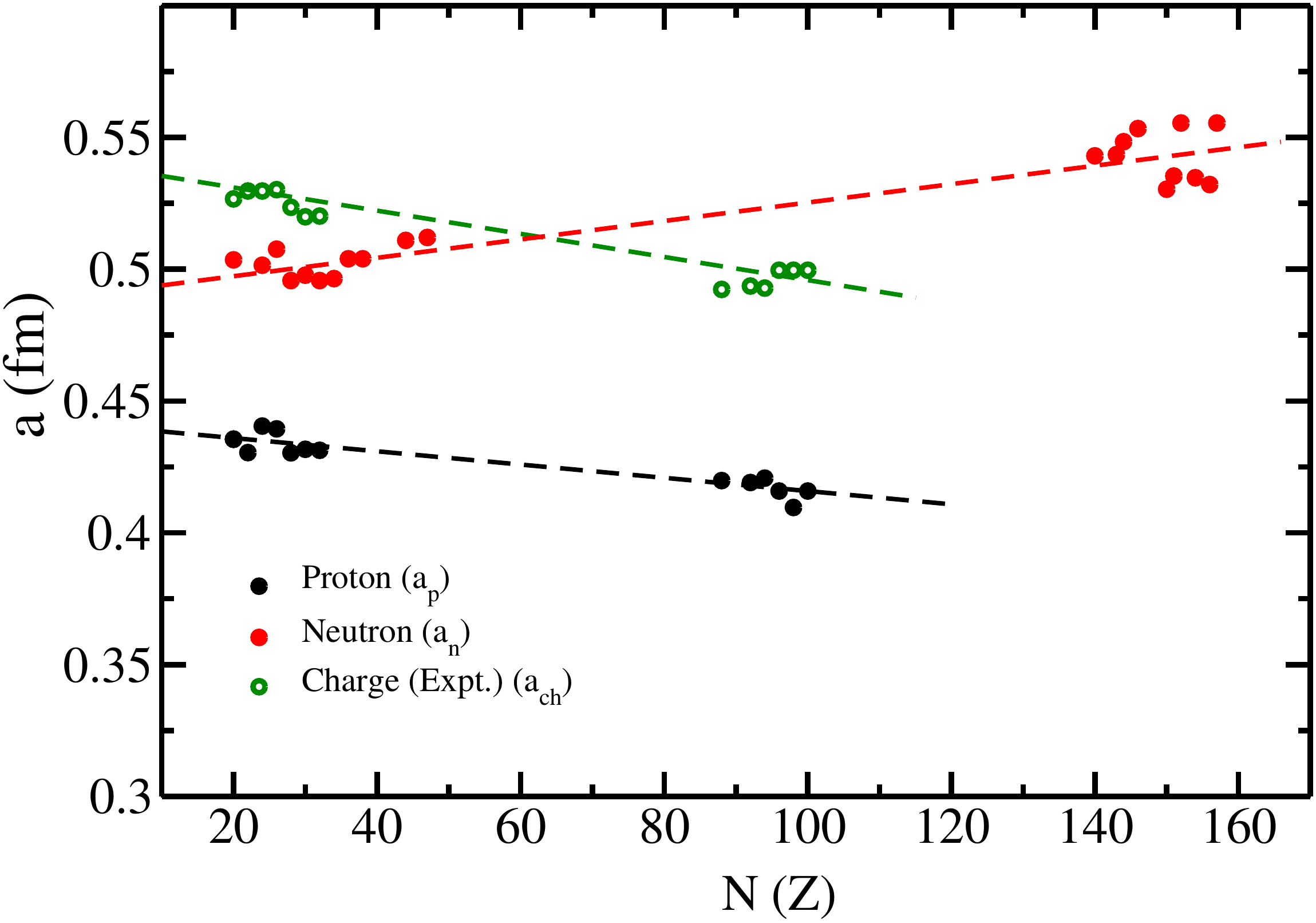}
\caption{(Color online) The surface diffusion parameter for all the targets and projectiles involved in the fusion cross section of superheavy isotopes of $Z$ = 120. The experimental data obtained for the charge radius are taken from Ref. \cite{angeli13}.}
 \label{fig8}
 \end{figure}
\begin{figure*}
\includegraphics[scale=.45]{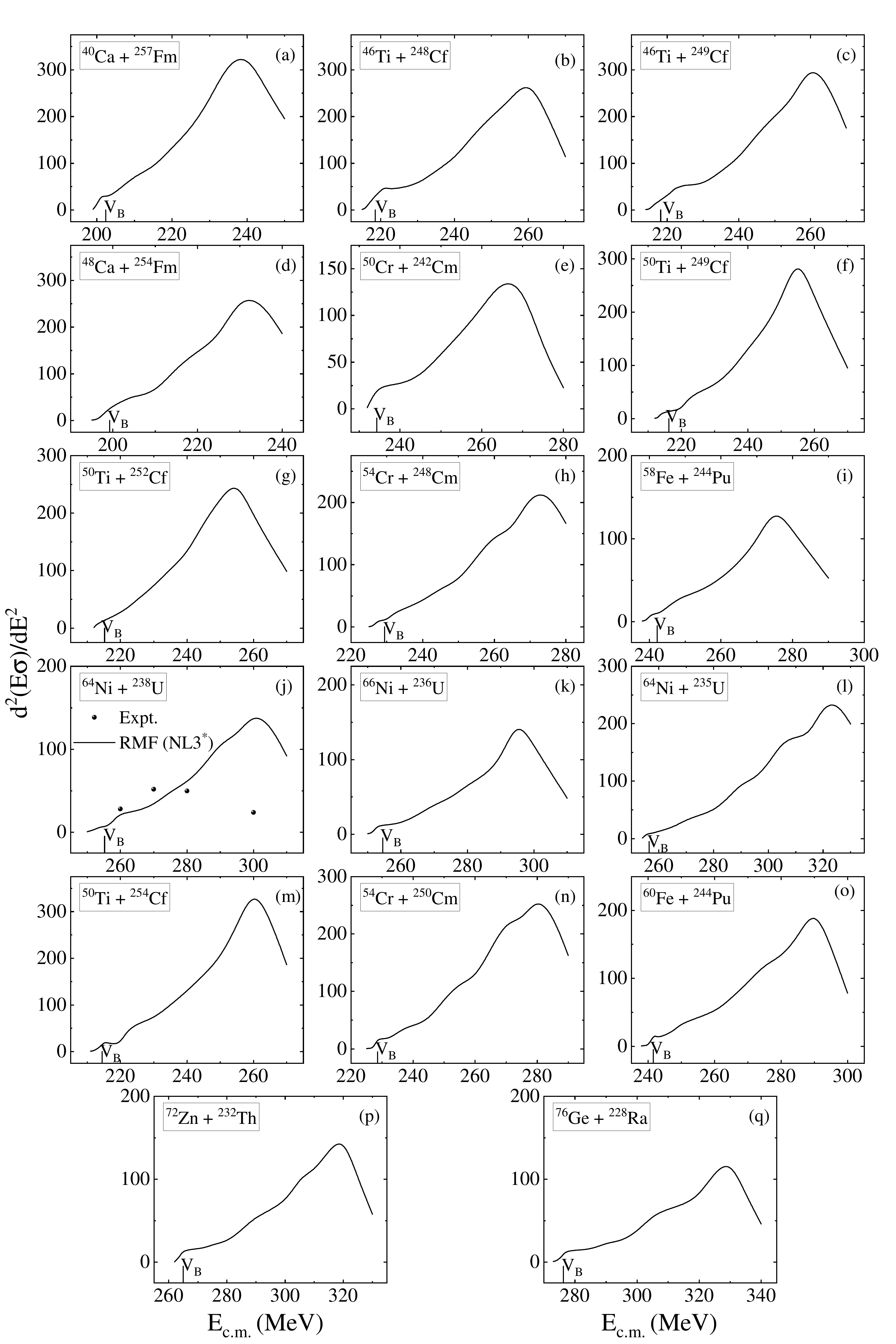}
\vspace{-0.5cm}
\caption{The barrier distributions for all the target-projectile combinations forming the SHN $Z$ = 120. The experimental data for reaction $^{64}$Ni + $^{238}$U are taken from Ref. \cite{kozu10}}
\label{fig9}
\end{figure*}


\subsection{Capture cross section} 
\noindent
The barrier height, position, and frequency are crucial requirements to calculate the capture and/or fusion cross-section for the two interacting nuclei, which can be extracted from the total interaction potential. From the results obtained in Refs. \cite{bhuy18,kuma09}, it is observed that the $\ell$-summed Wong formula given by Eq. (\ref{crs}) gives comparatively better agreement with the experimental data as compared to the simplified Wong formula. Hence, in the present study, the fusion cross section for the various isotopes of superheavy $Z$ = 120 is obtained using the $\ell$-summed Wong formula \cite{kuma09,bhuy18}.  We have calculated and compared capture cross sections for all the 17 projectile-target systems, i.e., $^{40}$Ca + $^{257}$Fm, $^{48}$Ca + $^{254}$Fm, $^{46}$Ti + $^{248}$Cf, $^{46}$Ti + $^{249}$Cf, $^{50}$Ti + $^{249}$Cf, $^{50}$Ti + $^{252}$Cf, $^{50}$Cr + $^{242}$Cm, $^{54}$Cr + $^{248}$Cm, $^{58}$Fe + $^{244}$Pu, $^{64}$Ni + $^{238}$U, $^{64}$Ni + $^{235}$U, $^{66}$Ni + $^{236}$U, $^{50}$Ti + $^{254}$Cf, $^{54}$Cr + $^{250}$Cm, $^{60}$Fe + $^{244}$Pu, $^{72}$Zn + $^{232}$Th, and $^{76}$Ge + $^{228}$Ra. In order to predict the suitable target-projectile combination (s) for the synthesis of a yet unknown atomic nucleus with $Z$ = 120, seven different isotopes, viz., $^{292}120$ ($N$ = 172), $^{294}120$ ($N$ = 174) , $^{295}120$ ($N$ = 175), $^{297}120$ ($N$ = 177), $^{299}120$ ($N$ = 179), $^{302}120$ ($N$ = 182), and $^{304}120$ ($N$ = 184) are considered here. Since the stability of superheavy nuclei is situated towards the neutron-rich side of the superheavy island \cite{ogan00,bhuy12,sama21,zhang05}, more neutron-rich target-projectile combinations are chosen here for the synthesis of $^{304}120$ ($N$ = 184),  predicted to be the next double magic nucleus beyond $^{208}$Pb \cite{bhuy12,agbe15,shi19,rutz97,zhang05}, and its neighboring isotope $^{302}120$ ($N$ = 182).

In Fig. \ref{fig3}, we show the capture cross section as a function of center-of-mass energy ($E_{c.m.}$) for all the 17 target-projectile systems. The experimental cross-section for the reaction $^{64}$Ni + $^{238}$U taken from Ref. \cite{kozu10} is also given for comparison. The  $\ell_{max}$ values for this system are calculated using the sharp cut off model \cite{beck81}. An interpolating polynomial is obtained for the $\ell_{max}$ values depending on the ratio $E_{c.m.}/V_B$ for the $^{64}$Ni + $^{238}$U system. This polynomial is further employed for extracting the $\ell_{max}$ values for the other reactions using their respective center-of-mass energies and barrier heights. The height of the fusion barrier ($V_B$) is also indicated on the $E_{c.m.}$ axis for all the reactions. From the figure, one can notice that that calculated cross-section (solid black line) shows an overall overlap with the experimental data (black spheres). Our previous study examined a few systems for the known region of the superheavy nuclei and found good predictions for the fusion characteristics using the  R3Y NN potential via the $\ell$-summed Wong formula \cite{bhuy20}. Hence, the $\ell-$summed Wong formula and the R3Y interaction potential can give reliable predictions of the fusion cross section for the unknown region of the superheavy island, which is crucial information for the future experimental synthesis of superheavy nuclei. Comparing all the cross sections obtained for the suitable system combination (projectiles and targets) for the synthesis of various isotopes of $Z$ =120, we notice a slightly large cross section for $^{304}$120 corresponding to the highly discussed next magic neutron $N$ =184 \cite{bhuy12,agbe15,rutz97,sobi66,nils69,shi19,zhang05}. 

To predict the appropriate target-projectile combination for the synthesis of a superheavy element $Z$ = 120, the capture cross sections for all the considered systems need to be compared at all the center-of-mass energy ($E_{c.m.}$) regions. To better understand all the reactions considered, we have normalized their $E_{c.m.}$'s for the respective Coulomb barriers. In Fig. \ref{fig4}, the capture cross section ($\sigma$) is displayed as a function of  $E_{c.m.}$ divided by estimated fusion barrier $V_B$ for all the target-projectile systems. In the below-barrier energy region ($E_{c.m.}/V_B < 1$), the reaction $^{50}$Ti + $^{249}$Cf is observed to yield the highest cross section among the considered systems. We also find a compatible capture cross section for the $^{40}$Ca + $^{257}$Fm system at energies above and/or around the Coulomb barrier. Although a large fusion/capture cross-section was obtained for the $^{40}$Ca + $^{257}$Fm system, the target $^{257}$Fm has a half-life of 100.5 days and also is available in a limited quantity of the order of picograms \cite{hoff03}, which do not allow for the production of a thick target of $^{257}$Fm at present \cite{robe15,hoff03}. Hence, a more feasible way to synthesize $Z$ = 120 is to move on towards the heavier projectiles than the Ca beam. 

Analogous to the structural prediction of Refs. \cite{bhuy12,agbe15,lala96,rutz97,sobi66,nils69,shi19,zhang05}, here we analyze the cross section at below- and/or above-barrier energies for the reaction systems associated with the compound nucleus of neutron number $N$ = 184  and its neighboring isotopes of $N$ =182 for $Z$ = 120. For example, in the case of $^{304}$120 ($N$ = 184), five reaction systems, $^{50}$Ti + $^{254}$Cf, $^{54}$Cr +
$^{250}$Cm, $^{60}$Fe + $^{244}$Pu, $^{72}$Zn + $^{232}$Th, and $^{76}$Ge + $^{228}$Ra are considered. Among these systems, $^{72}$Zn + $^{232}$Th and $^{50}$Ti + $^{254}$Cf are found to provide large capture cross sections at below- and above-barrier center-of-mass energies ($E_{c.m.}$). Here, $^{254}$Cf has a half-life of 60.5 $\pm$ 0.2 days \cite{phil63}, and so can be treated as a poor candidate for experimental synthesis. Hence the next prominent case is $^{54}$Cr + $^{250}$Cm which gives higher yield around the Coulomb barrier and can be considered as a suitable combination for the synthesis of $^{304}$120 ($N$ = 184). Similarly, in the case of $^{302}$120 ($N$ = 182), six reaction systems, $^{48}$Ca + $^{254}$Fm, $^{50}$Ti + $^{252}$Cf, $^{54}$Cr + $^{248}$Cm, $^{58}$Fe + $^{244}$Pu, $^{64}$Ni + $^{238}$U, and $^{66}$Ni + $^{236}$U, are considered. The highest capture cross section is observed for the $^{48}$Ca + $^{254}$Fm reaction among these six system combinations. The nucleus $^{254}$Fm has half-life of 3.24 hours \cite{akov01}, hence cannot be used as a target in the experiment. Further, at below-Coulomb-barrier energies, the reaction $^{66}$Ni + $^{236}$U is observed to give a higher capture cross-section but, again, $^{66}$Ni has a half-life of 56 hours \cite{john56}. The next two combinations, namely, $^{54}$Cr + $^{248}$Cm and $^{50}$Ti + $^{252}$Cf, are suitable for the synthesis of isotope $^{302}120$ at energies below and above the Coulomb barrier. Further, the structure of the interacting nuclei has a significant impact on the fusion cross section. In order to compare the cross sections for these reactions under study, the reduced cross sections are obtained, which exclude the structure effects. In terms of the collision radius, the reduced cross-section can be expressed as, $\sigma_{red}= \sigma/(A_{1}^{1/3}+A_{2}^{1/3})^2$. Figure \ref{fig5} shows the reduced cross section as a function of the center-of-mass energy divided by the height of the barrier (E$_{c.m.}$/V$_B$). The cross-section trend for all the systems is observed to be the same as in Fig. \ref{fig4}, which adheres to the predictions of capture cross-section.

\noindent
{\bf $Q$ values of the reactions:} The $Q$ value of a reaction depends upon the binding energies ($Q_{IN}= B_f-B_i$), where $B_f$ and $B_i$ are the sums of binding energies of products and reactants, respectively \cite{kran88} of the interacting nuclei and have significant effects upon the reaction characteristics. To account for this effect, the capture cross-section is displayed as function normalized excitation energy ($E^*=E_{c.m.} + Q_{IN}$) in Fig. \ref{fig6}. The $Q$ values are also calculated using the binding energy of projectiles and targets obtained from the axially deformed relativistic mean-field formalism for the NL3$^*$ parameter set. The normalization of E$^*$ is done for the sake of a comparison of different systems. Here V$^*$ (in MeV) is given as the sum of the $Q$ value and the observed Coulomb barrier of respective reactions.  It is observed from the figure that the capture cross sections of all the reactions converge near $E^{*}/V^{*}$=1. The reactions $^{66}$Ni + $^{236}$U and $^{40}$Ca + $^{257}$Fm are observed to give the highest cross section at below- and above-barrier excitation energies, respectively.  As mentioned above, $^{257}$Fm is not suitable to be used as a target due to its half-life \cite{hoff03,robe15}. Hence we have to consider the next best system at the above-barrier energies, which is $^{46}$Ti + $^{249}$Cf. Also eliminating reactions $^{66}$Ni + $^{236}$U and $^{48}$Ca + $^{254}$Fm due to shorter half-lives of $^{66}$Ni and $^{254}$Fm \cite{akov01,john56}, it is observed that the reactions $^{58}$Fe + $^{244}$Pu and $^{54}$Cr + $^{248}$Cm give the highest cross sections at below- and above-barrier excitation energies, respectively for isotope $^{302}120$.  At above barrier excitation energies the reaction $^{54}$Cr + $^{248}$Cm gives the highest cross section for isotope $^{302}$120. \\
Similarly, we have taken five different reactions for $^{304}120$ because it is predicted to be the next neutron magic ($N$ = 184) nucleus in the superheavy valley \cite{bhuy12,agbe15,rutz97,sobi66,nils69,shi19,zhang05}. Comparing the cross sections leading to formation of isotopes $^{304}120$, it is observed that the reactions $^{72}$Zn + $^{232}$Th and $^{50}$Ti + $^{254}$Cf give the highest cross sections at below- and above-barrier excitation energies, respectively. Again eliminating reaction $^{50}$Ti + $^{254}$Cf due to the 60.5 $\pm$ 0.2 days half-life of  $^{254}$Cf \cite{phil63}, the reaction $^{54}$Cr + $^{250}$Cm is found to yield the maximum cross section at above-barrier excitation energies. More detailed studies are highly welcome in correlation with the structural properties of these nuclei.   

\noindent
\textbf{Fusion cross section:} In the case of superheavy nuclei, the fusion cross section is obtained using the analytical expression given in Eq. (\ref{fus}). Figure \ref{fig7} represents fusion cross section ($\sigma_{fus}$) as a function of $E_{c.m.}$ divided by estimated fusion barrier $V_B$ for all target-projectile systems. From the figure, it is again noticed that the reaction $^{40}$Ca + $^{257}$Fm  has a large fusion cross section for the whole range of center-of-mass energies. Considering the isotope $^{302}120$, the reaction $^{48}$Ca + $^{254}$Fm is found to yield the maximum fusion cross section. However $^{257}$Fm and $^{254}$Fm cannot be used as a targets due to their limited quantities and shorter half-lives as mentioned above \cite{hoff03,robe15,akov01}. After $^{40}$Ca + $^{257}$Fm and $^{48}$Ca + $^{254}$Fm, the system $^{50}$Ti + $^{252}$Cf is found to give yield the highest fusion cross-section for the synthesis of SHN $^{302}120$. In the case of isotope  $^{304}120$, the reaction $^{50}$Ti+$^{254}$Cf is observed to yield the highest fusion cross-section. Since $^{254}$Cf cannot be used as a target nuclei because of its shorter half-life of 60.5 $\pm$ 0.2 days \cite{phil63}, we have to consider the next reaction system which is observed to be $^{54}$Cr+$^{250}$Cm. Further, the fusion cross section decreases with the increase in the atomic number of the projectile as can be noted in Fig. \ref{fig7}. From the above discussion it is inferred that the isotopes of  $^{x}$Ti + $^y$Cf are the most suitable projectile-target combination for the synthesis of SHN $Z$ = 120. However in case of isotope $^{304}120$, the reaction $^{54}$Cr + $^{250}$Cm is found to be more suitable.

\noindent
{\bf Surface diffuseness parameter:} The nuclear surface diffuseness parameter is calculated for all the interacting nuclei to connect the surface properties in terms of nuclear density distributions to the fusion cross section. The equivalent nuclear surface diffuseness parameter can be obtained using the relation $a_i \approx -\rho_i/ \frac{d\rho_i}{dr}$, where $i$ stands for proton- ($a_p$), neutron- ($a_n$) and charge ($a_{ch}$) density distributions of the nucleus. The diffusion parameter results for proton and neutron density distribution from the relativistic mean-field approach using the NL3$^*$ parameter along with the experimental charge density \cite{vrie87,nadj94,angeli13} are displayed in Fig. \ref{fig8}. The black solid, red solid, and green open circles in the figure represent equivalent nuclear surface diffuseness parameters for proton ($a_p$), neutron ($a_n$) and charge ($a_{ch}$) density, respectively.  The experimental charge density is extracted from electron scattering \cite{vrie87,nadj94,angeli13}.  The figure shows that the magnitude of the equivalent surface diffusion parameter for experimental charge densities is greater than the proton density from our calculation. This difference ($\approx 0.1$ $fm$) is due to the finite size effect in the proton density, which is not considered in the RMF calculation. The proton density is called charge density in experimental data due to the charge effect, similar to the proton radius and charge radius. On connecting the cross section with the equivalent nuclear surface diffuseness parameters for the proton and neutron corresponding to the target and/or projectiles combinations, a poor correlation can be established among them. For example, a large cross section is obtained for the combination of the projectile and/or target having a relatively small magnitude of equivalent surface diffuseness parameter. More systematic studies in this direction are highly welcome. \\ 

\noindent
{\bf Barrier distribution:} As discussed above, the potential barrier is formed from the long-ranged repulsive Coulomb force among the protons and the short-ranged attractive nuclear force among the nucleons. Rowley {\it et al.} \cite {rowl91} described a method to obtain the smoothed barrier distribution from the measured cross section at near- and sub-barrier energies. The fusion barrier distribution function ($\frac{d^2(E.\sigma)}{dE^2}$) is obtained by double differentiation of the transmission function ($E.\sigma$) with respect to center-of-mass energy. Figure \ref{fig9} shows the fusion barrier distribution for all the target-projectile combinations. For the reaction, $^{64}$Ni + $^{238}$U the experimental barrier distribution (solid black circles) is also obtained using data from Ref. \cite{kozu10}.  The cross-section experimental values are available only at four center-of-mass energies, which lead to a difference between the theoretical and experimental values of barrier distributions. The value of calculated fusion barrier height is also indicated for each reaction. The highest barrier height is observed for the reaction  $^{50}$Ti + $^{254}$Cf whereas the lowest barrier height is observed for the $^{76}$Ge + $^{228}$Ra reaction. \\
\begin{table}
\caption{\label{tab1} The ground state quadrupole deformation ($\beta_2$) for the projectiles and targets within relativistic mean field using non-linear NL3$^*$ parameter sets. Finite range droplet model (FRDM) \cite{moll16} predictions and the available experimental data \cite{raman01,prit12} are given for comparison.}
\renewcommand{\tabcolsep}{0.29cm}
\renewcommand{\arraystretch}{1.2}
\begin{tabular}{cccccc} 
\hline
\multirow{2}{*}{Nuclei} & \multicolumn{3}{c}{Quadrupole deformation ($\beta_2$)}     \\ 
\cline{2-4}
& RMF (NL3$^*$) & FRDM    & Expt. \\
\hline
$^{40}$Ca      & 0.007       & 0.000            & 0.1230  $\pm$ 0.0110  \\ 
$^{48}$Ca      & 0.011       & 0.000            & 0.1060 $\pm$ 0.0180 \\
$^{46}$Ti      & 0.013       & 0.021            & 0.3170 $\pm$ 0.0080  \\
$^{50}$Ti      & 0.005       & 0.000            & 0.1660 $\pm$ 0.0110  \\ 
$^{50}$Cr      & 0.223       & 0.194            & 0.2912 $\pm$ 0.0032 \\
$^{54}$Cr      & 0.165       & 0.161             & 0.2509$\pm$ 0.0075  \\ 
$^{58}$Fe      & 0.209       & 0.173             & 0.2610 $\pm$ 0.0050 \\
$^{60}$Fe      & 0.210      & 0.185             & 0.2240 $\pm$ 0.0100 \\
$^{64}$Ni      & 0.078      &-0.094             & 0.1628 $\pm$ 0.0041 \\
$^{66}$Ni      & 0.039      & 0.000             & 0.1570 $\pm$  0.0090 \\
$^{72}$Zn      & 0.013      & 0.011             & 0.2340 $\pm$  0.0014 \\
$^{76}$Ge      & 0.171      & 0.161             & 0.2623 $\pm$ 0.0039  \\
$^{228}$Ra     & 0.213      & 0.174             & 0.2170 $\pm$ 0.0050 \\
$^{232}$Th     & 0.251      & 0.205             & 0.2608 $\pm$ 0.0014  \\
$^{235}$U      & 0.269      & 0.215             &   -                 \\
$^{236}$U      & 0.275      & 0.226             & 0.2821 $\pm$ 0.0018 \\
$^{238}$U      & 0.283      & 0.236             & 0.2863 $\pm$ 0.0024  \\
$^{244}$Pu     & 0.296      & 0.237             & 0.2931 $\pm$ 0.0017 \\
$^{242}$Cm     & 0.292      & 0.237             &  -                   \\
$^{248}$Cm     & 0.295      & 0.250             & 0.2983 $\pm$ 0.0019  \\
$^{250}$Cm     & 0.289      & 0.250             & 0.2972 $\pm$ 0.0019 \\
$^{248}$Cf     & 0.297      & 0.250             &  -                  \\
$^{249}$Cf     & 0.297      & 0.250             &  -                   \\
$^{252}$Cf     & 0.296      & 0.251             & 0.3040 $\pm$ 0.0100 \\
$^{254}$Cf     & 0.285      & 0.240             &  -                   \\
$^{254}$Fm     & 0.294      & 0.251             & -                   \\
$^{257}$Fm     & 0.276      & 0.241             & -                    \\
\hline           
\end{tabular}
\end{table}
\begin{figure}
\centering
\includegraphics[scale=0.35]{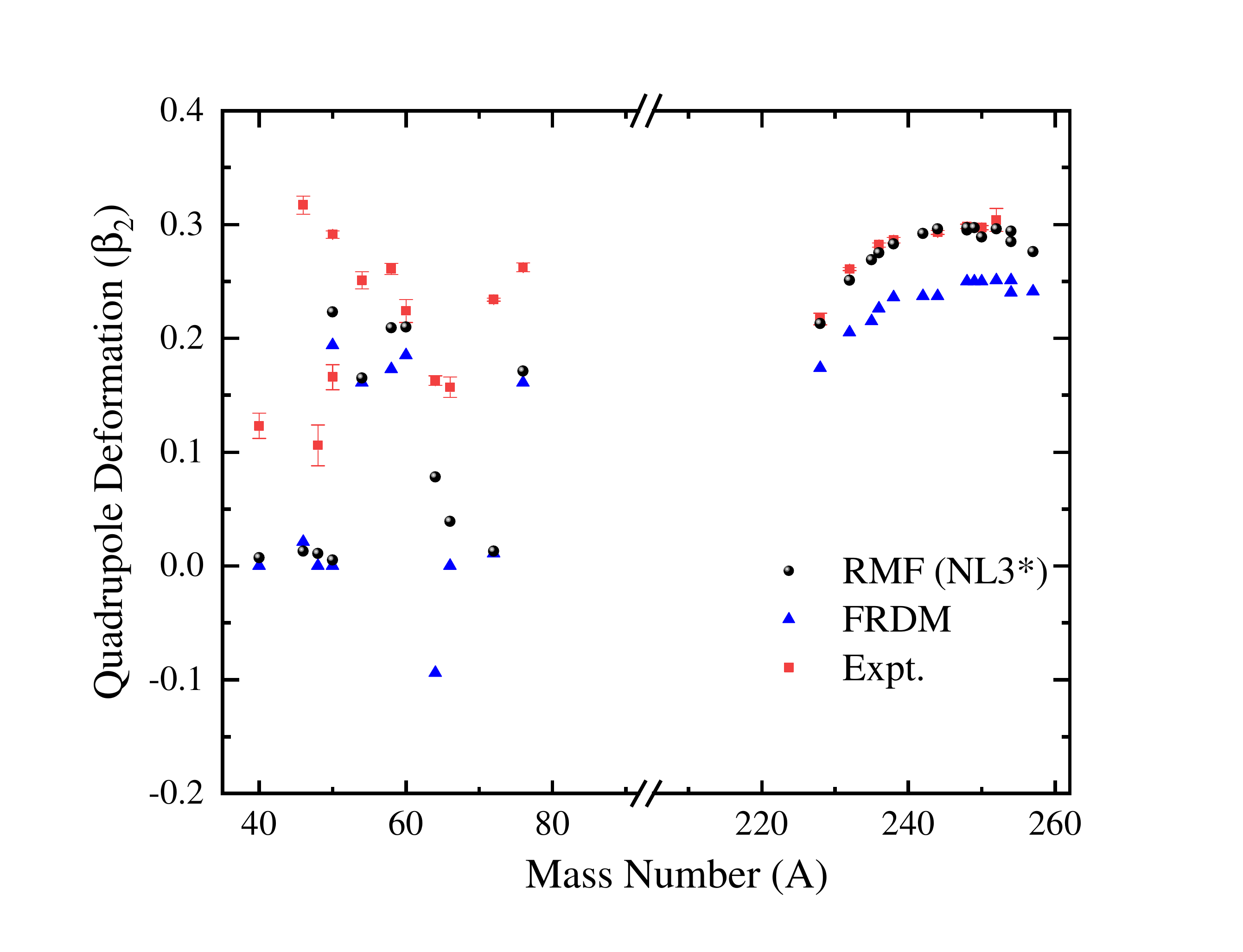}
\caption{(Color online) The ground state quadrupole deformation parameter ($\beta_2$) plotted as a function of mass number ($A$) of all the interacting target and projectile nuclei.}
\label{beta2}
\end{figure}

\noindent
{\bf Nuclear deformations and cross section:}
We have taken various projectile and target combinations in the present analysis to predict the possible candidates for synthesizing superheavy nuclei. Due to the model's present limitation, we have taken the spherical density distributions to generate the nuclear interaction potential. Although the nuclear potential is generated from spherical density distributions of the projectiles and targets, the energy, i.e., the $Q-$value of the reaction, is obtained from the binding energy corresponding to the axially deformed ground state using the relativistic mean-field model for the NL3$^*$ parameter sets. To justify the above results, we analyze the ground state deformations for these considered nuclei. The quadrupole deformation parameter ($\beta_2$) for targets and projectiles in their ground state are obtained from the axially deformed relativistic mean-field formalism. The values for NL3$^*$ parameter sets (black circle) are listed in Table \ref{tab1}, and also shown in Fig. \ref{beta2} along with the finite range droplet model (FRDM) predictions  (blue triangles) \cite{moll16} and the experimental data (red square) \cite{raman01,prit12}.  A comprehensive agreement can be observed from Table \ref{tab1} as well as from Fig. \ref{beta2} for RMF (NL3$^*$) with experimental data and FRDM predictions. \\
One can notice from the figure that the $\beta_2$ values of all the target nuclei are of the same order ($\approx 0.25$), whereas there is variation in the $\beta_2$ values for the projectiles. In other words, all the target nuclei are prolate in their ground state, but the projectile attains a spherical or prolate shape. As we know, a positive value of $\beta_2$, i.e., a prolate target and/or projectile nucleus, increases the fusion cross section at the sub-barrier center-of-mass energies as compared to the spherical one \cite{sarg11}. Since the value of $\beta_2$ for all the considered target nuclei is of the same order and the projectiles of the feasible, competing, and favorable reactions also have the same order of $\beta_2$, it will increase the capture cross-section in the same order of magnitude at sub-barrier energies for these reaction systems \cite{sarg11}. Hence, we may assume that nuclear deformation will change the magnitude of the cross-section but hardly affect the conclusions of the present analysis, as here we are interested in the relative performance of the reactions. The implication of the shape degrees of freedom within relativistic mean-field for the study of fusion/capture cross-section is in progress.
   
\section{SUMMARY AND CONCLUSIONS} 
\label{summary}
\noindent
The capture and/or fusion cross sections for seventeen projectile target combinations, namely, $^{40}$Ca + $^{257}$Fm, $^{48}$Ca + $^{254}$Fm, $^{46}$Ti + $^{248}$Cf, $^{46}$Ti + $^{249}$Cf, $^{50}$Ti + $^{249}$Cf, $^{50}$Ti + $^{252}$Cf, $^{50}$Cr + $^{242}$Cm, $^{54}$Cr + $^{248}$Cm, $^{58}$Fe + $^{244}$Pu, $^{64}$Ni + $^{238}$U, $^{64}$Ni + $^{235}$U, $^{66}$Ni + $^{236}$U, $^{50}$Ti + $^{254}$Cf, $^{54}$Cr + $^{250}$Cm, $^{60}$Fe + $^{244}$Pu, $^{72}$Zn + $^{232}$Th, and $^{76}$Ge + $^{228}$Ra, have been calculated. These reactions lead to seven different neutron-rich isotopes, $^{292}120$ ($N$ = 172), $^{294}120$ ($N$ = 174), $^{295}120$ ($N$ = 175), $^{297}120$ ($N$ = 177), $^{299}120$ ($N$ = 179), $^{302}120$ ($N$ = 182) and $^{304}120$ ($N$ = 184), of the element $Z$ = 120.  The relativistic R3Y NN interaction potential and spherical density distributions from the relativistic mean-field (RMF) formalism are used to estimate the nuclear interaction potential by adopting a double folding procedure.  The capture cross section is estimated by using the well-known $\ell-$summed Wong formula. We find a reasonably good agreement among the calculated capture cross section with $\ell_{max}$ values from the sharp cut off model and experimental cross section for reaction $^{64}$Ni + $^{238}$U. We have shown the capture cross-section as a function of normalized center-of-mass energy with respect to fusion barrier for all the reaction systems. \\

We find that the reactions $^{50}$Ti + $^{249}$Cf and $^{40}$Ca + $^{257}$Fm give a large cross section in below- and/or above-barrier regions. The reaction $^{40}$Ca + $^{257}$Fm leads in fusion cross section at all energies. For the synthesis of $^{302}120$ and $^{304}120$, the reactions $^{48}$Ca + $^{254}$Fm and $^{50}$Ti + $^{254}$Cf are found to yield the maximum fusion cross section. $^{257}$Fm is not available in sufficient quantity \cite{hoff03,robe15}, whereas $^{254}$Fm and $^{254}$Cf have half-lives of 3.24 hours \cite{akov01} and 60.2 days \cite{phil63}, respectively. Hence, it is not feasible at the present time to use these isotopes as  targets experimentally. So we have to consider the next possible reactions, \emph{i.e.}, $^{50}$Ti + $^{252}$Cf  and $^{54}$Cr + $^{250}$Cm as the possible predicted candidates for the synthesis of $^{302}120$ and $^{304}120$, respectively. From the observations of cross section ($\sigma$) versus excitation-energy ($E^*$), we noticed that the reactions $^{66}$Ni + $^{236}$U and $^{48}$Ca + $^{254}$Fm provide large cross sections, while $^{66}$Ni has a half-life of 56 hours \cite{john56}. This is again complicated to use as a projectile in the laboratory. Based on the above analysis, we observed that isotopes of Ti + Cf are predicted to be the most suitable target-projectile combinations for the synthesis of superheavy of $Z$ = 120.  However, in the case of $^{304}$120, due to the shorter half-life of $^{254}$Cf, the reaction $^{54}$Cr + $^{250}$Cm is found to be a more suitable candidate.  Recent experimental observations have shown $^{50}$Ti + $^{249}$Cf to be the most promising reaction for the synthesis of SHN $Z$ = 120 \cite{albe20}.  In addition to these, we also correlate the equivalent surface diffuseness parameter with the reaction system in terms of nucleon density distributions of the projectiles and targets. We found a linear dependence of the surface diffuseness with the fusion cross section. In other words, a larger fusion/capture cross section of the projectile and/or target is associated with a smaller value of the neutron and/or proton equivalent surface diffuseness parameter. A more systematic study including shape degrees of freedom within the relativistic mean-field approach will be performed shortly.

\section*{Acknowledgments}
This work was supported by DAE-BRNS Project Sanction No. 58/14/12/2019-BRNS, FOSTECT Project Code: FOSTECT.2019B.04, and FAPESP Project No. 2017/05660-0.


\end{document}